\documentclass[twocolumn]{aastex631}
\usepackage{amsmath}
\usepackage{natbib}
\usepackage{graphicx}
\usepackage{url}

\usepackage{verbatim}

\shorttitle{Proplyds with ALMA Band 3}
\shortauthors{Ballering et al.}

\begin{document}

\title{Isolating Dust and Free-Free Emission in ONC Proplyds with ALMA Band 3 Observations}

\author[0000-0002-4276-3730]{Nicholas P. Ballering}
\altaffiliation{Virginia Initiative on Cosmic Origins Fellow.}
\affiliation{Department of Astronomy, University of Virginia, Charlottesville, VA 22904, USA}

\author[0000-0003-2076-8001]{L. Ilsedore Cleeves}
\affiliation{Department of Astronomy, University of Virginia, Charlottesville, VA 22904, USA}
\affiliation{Department of Chemistry, University of Virginia, Charlottesville, VA 22904, USA}

\author[0000-0002-9593-7618]{Thomas J. Haworth}
\affiliation{Astronomy Unit, School of Physics and Astronomy, Queen Mary University of London, London E1 4NS, UK}

\author[0000-0001-8135-6612]{John Bally}
\affiliation{Center for Astrophysics \& Space Astronomy, Astrophysical \& Planetary Sciences Department, University of Colorado, UCB 389, Boulder, CO 80309, USA}

\author[0000-0002-1031-4199]{Josh A. Eisner}
\affiliation{Steward Observatory, University of Arizona, 933 N. Cherry Ave., Tucson, AZ 85719, USA}

\author[0000-0001-6431-9633]{Adam Ginsburg}
\affiliation{Department of Astronomy, University of Florida, P.O. Box 112055, Gainesville, FL, USA}

\author[0000-0001-9857-1853]{Ryan D. Boyden}
\affiliation{Steward Observatory, University of Arizona, 933 N. Cherry Ave., Tucson, AZ 85719, USA}

\author[0000-0001-8060-1321]{Min Fang}
\affiliation{Purple Mountain Observatory, Chinese Academy of Sciences, 10 Yuanhua Road, Nanjing 210023, People's Republic of China}

\author[0000-0001-6072-9344]{Jinyoung Serena Kim}
\affiliation{Steward Observatory, University of Arizona, 933 N. Cherry Avenue., Tucson, AZ 85719, USA}

\correspondingauthor{Nicholas P. Ballering}
\email{nb2ke@virginia.edu}

\begin{abstract}
The Orion Nebula Cluster (ONC) hosts protoplanetary disks experiencing external photoevaporation by the cluster's intense UV field. These ``proplyds" are comprised of a disk surrounded by an ionization front. We present ALMA Band 3 (3.1 mm) continuum observations of 12 proplyds. Thermal emission from the dust disks and free-free emission from the ionization fronts are both detected, and the high-resolution (0$\farcs$057) of the observations allows us to spatially isolate these two components. The morphology is unique compared to images at shorter (sub)millimeter wavelengths, which only detect the disks, and images at longer centimeter wavelengths, which only detect the ionization fronts. The disks are small ($r_d$ = 6.4--38 au), likely due to truncation by ongoing photoevaporation. They have low spectral indices ($\alpha \lesssim 2.1$) measured between Bands 7 and 3, suggesting the dust emission is optically thick. They harbor tens of Earth masses of dust as computed from the millimeter flux using the standard method, although their true masses may be larger due to the high optical depth. We derive their photoevaporative mass-loss rates in two ways: first, by invoking ionization equilibrium, and second using the brightness of the free-free emission to compute the density of the outflow. We find decent agreement between these measurements and $\dot M$ = 0.6--18.4 $\times$ 10$^{-7}$ $M_\odot$ yr$^{-1}$. The photoevaporation timescales are generally shorter than the $\sim$1 Myr age of the ONC, underscoring the known ``proplyd lifetime problem." Disk masses that are underestimated due to being optically thick remains one explanation to ease this discrepancy.
\end{abstract}


\section{Introduction}
\label{sec:introduction}

The Orion Nebula Cluster (ONC) provides an ideal site to study star and planet formation in a cluster environment. Clusters are where most stars and planets form \citep{lada2003}, and evidence suggests our solar system formed in a cluster \citep{adams2010}. The ONC comprises thousands of 1--2 Myr old stars \citep{fang2021_ONCHR} and resides at a distance of $\sim$400 pc \citep{jeffries2007_ONCdistance,menten2007_ONCdistance,kounkel2017_oriondistances}. Unlike low-mass star-forming regions (e.g., Taurus-Auriga, $\rho$ Ophiuchus, Lupus, Chamaeleon, etc.), clusters host luminous OB stars, which irradiate their surroundings with UV photons. In the ONC, these are the Trapezium stars, most notably the O7V (binary) star $\theta^1$Ori C \citep{sota2011_Ostars}.

The high-UV environment is expected to heat and photoevaporate protoplanetary disks around young stars in the cluster, driving mass loss via a wind and truncating the disks' radii \citep{johnstone1998_photoevap,storzer1999_proplyds,adams2004_photoevap,clarke2007_photoevaporation,winter2018_truncation,sellek2020_dustphotoevap,marchington2022_diskradii,winter2022_photoevapreview}. This impacts the physical and chemical properties of the disks \citep{walsh2013_irradiationchem,boyden2023_ONCgas} and the planet formation process occurring within \citep{throop2005_evapplanetesimals,haworth2018_trappist,winter2022_growth+migration,qiao2023_pebbles}. Observations of the ONC with the Atacama Large Millimeter/submillimeter Array (ALMA) found a deficit of large disks compared with low-mass star-forming regions, providing evidence for disk truncation by photoevaporation \citep{eisner2018_ALMAB7,boyden2020_ONCgas,otter2021_OMC,boyden2023_ONCgas}. However, other mechanisms, such as dynamical encounters and interstellar medium (ISM) interactions \citep{wijnen2017_truncation} or higher cosmic-ray ionization \citep{kuffmeier2020_CRdisksizes}, could also explain the small disk sizes in clusters like the ONC.

Direct evidence for ongoing external photoevaporation is provided by the proplyds---disks surrounded by an ionization front with a comet-like morphology. The external UV field heats the disks, driving material off in a photoevaporative wind. In most observed proplyds, the wind is neutral and driven by far-UV (FUV), rather than extreme-UV (EUV), photons. The outflowing material eventually encounters ionizing photons, mostly from $\theta^1$Ori C, yielding the observable ionization front that is well separated from the outer edge of the disk. Proplyds have recently been discovered in other star-forming regions \citep{kim2016_NGC1977proplyds,haworth2021_flameproplyds} although the ONC remains the site of the majority of known proplyds.

The morphology of proplyds was first observed in Hubble Space Telescope (HST) images of the ONC by \citet{odell1993_HSTONC}, and proplyds have been targeted with numerous HST observations since \citep{odell1994_HSTONC,odell1996_ONCHST,bally1998_ONCHST,bally2000_ONC,ricci2008_ONCHST}.

At centimeter wavelengths, the free-free emission from proplyd ionization fronts was detected with the Very Large Array \citep[VLA;][]{churchwell1987_VLAONC}. While the sources were slightly extended, their morphology was not clearly visible in these early radio observations. More recent observations with the VLA did reveal the structure of many ONC proplyds at centimeter wavelengths \citep{forbrich2016_JVLAONC,sheehan2016_ONCVLA}, although neither of these studies performed an analysis of proplyd morphology.

Early (sub)millimeter observations were only sensitive enough to detect the brightest ONC sources and lacked the spatial resolution to image proplyds in detail \citep{mundy1995_trapdisks,bally1998_OVRO,williams2005_proplydsSMA,eisner2006_trapezium}. ALMA offered a significant improvement in both sensitivity and resolution, with observations of the ONC at Band 6 \citep[1.3 mm;][]{eisner2016_ALMA6ONC} and Band 7 \citep[0.86 mm;][]{mann2014_ALMA7ONC,eisner2018_ALMAB7} detecting many sources, including proplyds. However, at these wavelengths the extended free-free emission from the ionization fronts was not seen; the observations instead detected compact sources arising from dust in the protoplanetary disks.

Here we present ALMA Band 3 (3.1 mm) observations of the ONC, focusing on 12 proplyd systems. At this wavelength, ALMA is sensitive to both dust emission from the disk and free-free emission from the ionization front. With a resolution of 0$\farcs$057, these observations can spatially resolve these two components, allowing each to be characterized independently.

\section{Methods}
\label{sec:methods}

\subsection{ALMA Band 3 Observations}

ALMA Band 3 observations of the ONC were performed in Cycle 6 project 2018.1.01107.S. The ONC was mosaicked with ten pointings, labeled Field IDs 5--14. Four observation blocks were carried out, the first (observation ID 0) on 2019 July 7 and the remaining three (observation IDs 1, 2, and 3) on 2019 July 9. The C43-9 configuration was used with 43 antennas in observations 0, 2, and 3, and 45 antennas in observation 1. Four spectral windows were used with central frequencies of 90.5, 92.4, 102.5, and 104.4 GHz. Each spectral window had a bandwidth of 1.875 GHz sampled with 1920 channels. Separate spectral window IDs were assigned for each of the four observations, resulting in 16 total spectral window IDs labeled 0--15.

We began our reduction with the pipeline-calibrated measurement set. Because we are only interested in the continuum, we flagged channels exhibiting line emission. This was done by inspecting the amplitude versus channel plot of each spectral window for each field. We then spectrally averaged the data with a bin width of 128 channels to produce a continuum measurement set with 15 channels per spectral window.

\begin{figure}
\epsscale{1.17}
\plotone{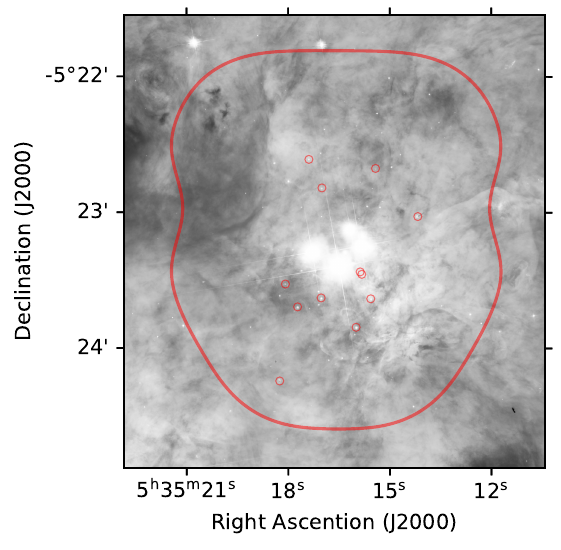}
\caption{HST/Advanced Camera for Surveys F435W image of the ONC center by \citet{robberto2013_ONCHST} overlaid with the field of view of the ALMA Band 3 mosaic and the locations of the 12 proplyds presented in this paper (red circles). The bright Trapezium stars are visible in the center of the image. North is up, and East is left.}
\label{fig:field}
\end{figure}

These observations detect many sources, necessitating analysis of the image plane rather than of the visibilities directly. We imaged the continuum with the CASA task tclean in multi-frequency synthesis (mfs) mode. All ten fields were imaged together using the mosiac gridder with a phase center at 05:35:16.578 $-$05:23:12.150. The pixel size is 0$\farcs$008 and the image size is 19,600 $\times$ 22,500 pixels (2$\farcm$6 $\times$ 3$\arcmin$), which includes a signal-free buffer region around the mosaic. The region of the mosaic with signal spans 05:35:11.71 to 05:35:21.45 in R.A. and $-$05:21:48.5 to $-$05:24:35.8 in decl. although it does not completely fill this rectangle, and the signal-to-noise ratio is poor near the edges. The footprint of the mosaic on an HST image is shown in Figure \ref{fig:field}. The mtmfs deconvolver was used with point-source model components and two spectral terms. Briggs weighting was used with a robust parameter of 0.5, resulting in a beam size of 0$\farcs$057 $\times$ 0$\farcs$072 (23 $\times$ 29 au at 400 pc). This resolution is comparable to the famous HST images of proplyds in the optical \citep[e.g., 0\farcs05;][]{bally1998_ONCHST}. We also imaged the mosaic a second time with a robust parameter of 2.0, yielding a beam size of 0$\farcs$069 $\times$ 0$\farcs$107 (28 $\times$ 43 au). This image has lower spatial resolution but higher sensitivity to extended emission.

Upon initially imaging the data from the four observations combined, we noticed persistent artifacts surrounding some of the brightest sources, most notably $\theta^1$ Ori A. We found the artifacts are due in part to source variability. Similar imaging artifacts were seen around $\theta^1$ Ori A in VLA observations \citep{forbrich2016_JVLAONC}. To address this, we first modeled the three brightest sources in the mosaic---$\theta^1$ Ori A, BN, and Source I---for each observation separately and used the CASA task uvsub to subtract this model from the data before imaging all observations together. This effectively removed these sources, and their artifacts, from the final image.

We self-calibrated the data during the initial per-observation imaging of the three bright sources. We derived and applied two iterations of phase calibration solutions for each field individually and thus to only the fields containing signal from these three sources. This resulted in modestly higher S/N in these fields.

The final imaging was performed with the automasking capabilities of the tclean task (using the ``auto-multithresh" algorithm) to generate clean masks around sources in the mosaic. After some experimentation, we found that a side-lobe threshold value of 1.5 and a noise threshold value of 4.5 worked well to identify and clean most observable sources while converging relatively quickly. Lastly, we corrected the mosaic for the response of the primary beam using the widebandpbcor task.

\subsection{Supplementary Data}

To obtain a multiwavelength picture of our sources, we also acquired the ALMA Band 7 (0.86 mm) image of the ONC published by \citet{eisner2018_ALMAB7} and the VLA images at $K$ (1.3 cm), $X$ (3.6 cm), and $C$ (6 cm) Bands published by \citet{sheehan2016_ONCVLA}.   

\subsection{Source Identification and Measurements}

\begin{figure*}
\gridline{\fig{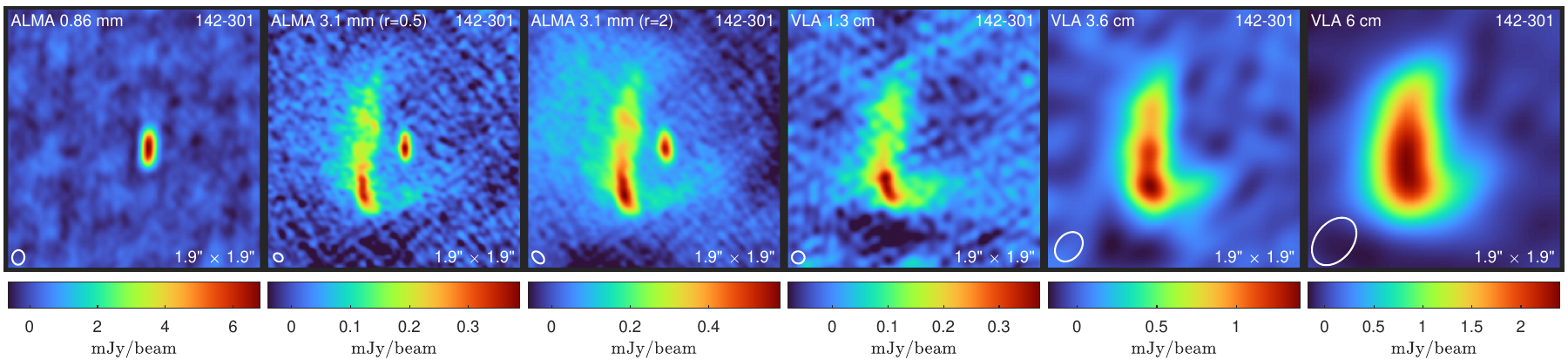}{1.0\textwidth}{}}\vspace{-10mm}
\gridline{\fig{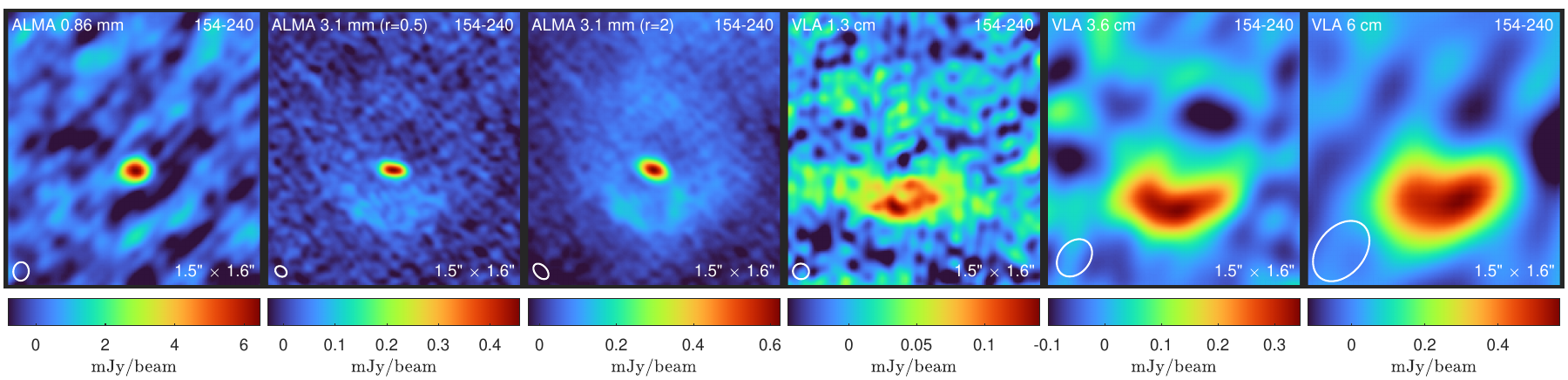}{1.0\textwidth}{}}\vspace{-10mm}
\gridline{\fig{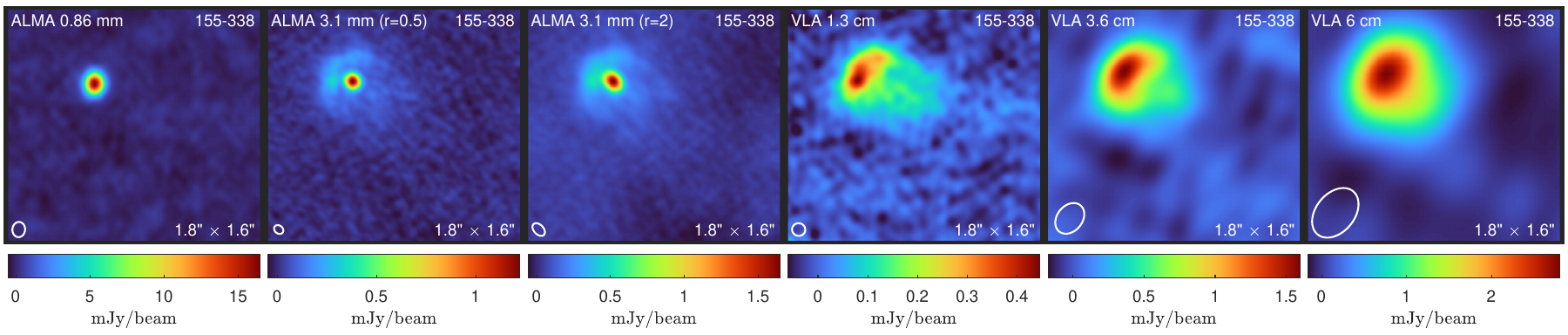}{1.0\textwidth}{}}\vspace{-10mm}
\gridline{\fig{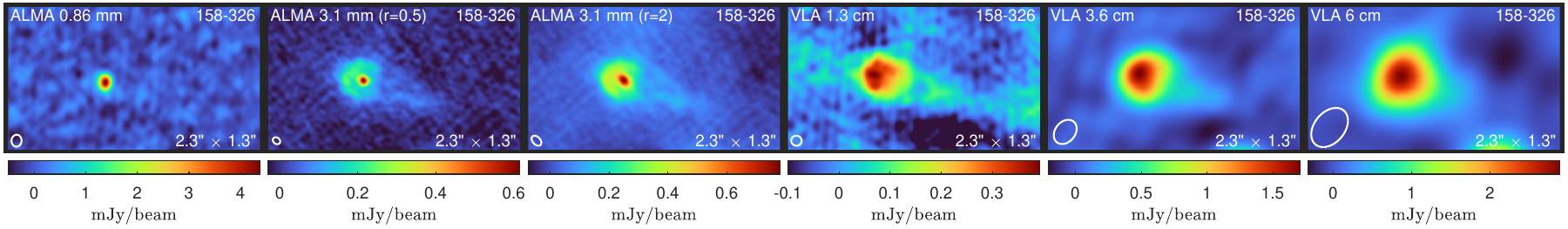}{1.0\textwidth}{}}\vspace{-10mm}
\gridline{\fig{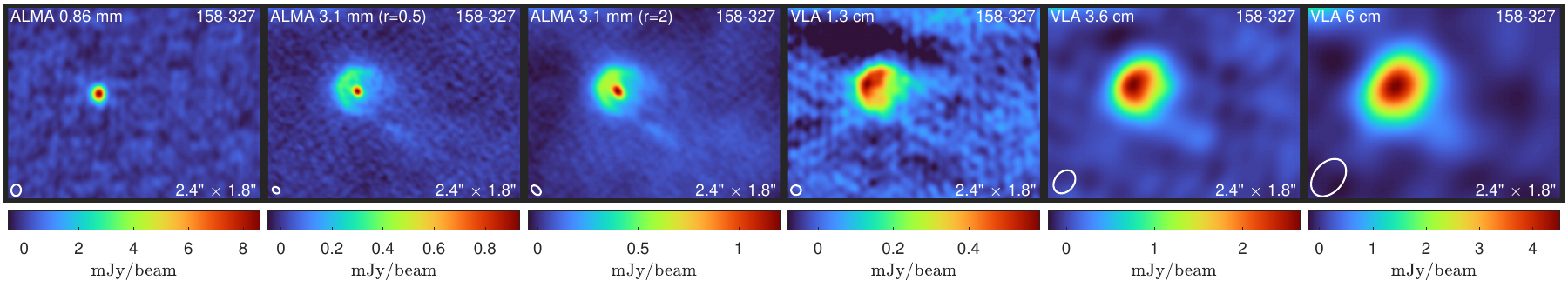}{1.0\textwidth}{}}
\caption{Images of proplyds 142-301, 154-240, 155-338, 158-326, 158-327, and 159-359 at five wavelengths. ALMA Band 7 (0.86 mm) images, originally published by \citet{eisner2018_ALMAB7}, are shown in the left column. Our ALMA Band 3 (3.1 mm) images CLEANed with a robust parameter of 0.5 and 2.0 are shown in the second and third columns, respectively. The fourth, fifth, and sixth columns show VLA images at $K$ Band (1.3 cm), $X$ Band (3.6 cm), and $C$ Band (6 cm) from \citet{sheehan2016_ONCVLA}. The size of each image is noted in the lower right, and the beam size is indicated by the white ellipse in the lower left. For scale, the beam size of the ALMA 3.1 mm (r=0.5) image is 23 $\times$ 29 au. The color represents the surface brightness in units of millijansky/beam. The ALMA Band 7 images detect thermal dust emission from the disk, whereas the VLA images detect free-free emission from the proplyd ionization fronts. The Band 3 images reveal emission from both the dust and free-free components.} \label{fig:proplyds1}
\end{figure*}

\begin{figure*}
\gridline{\fig{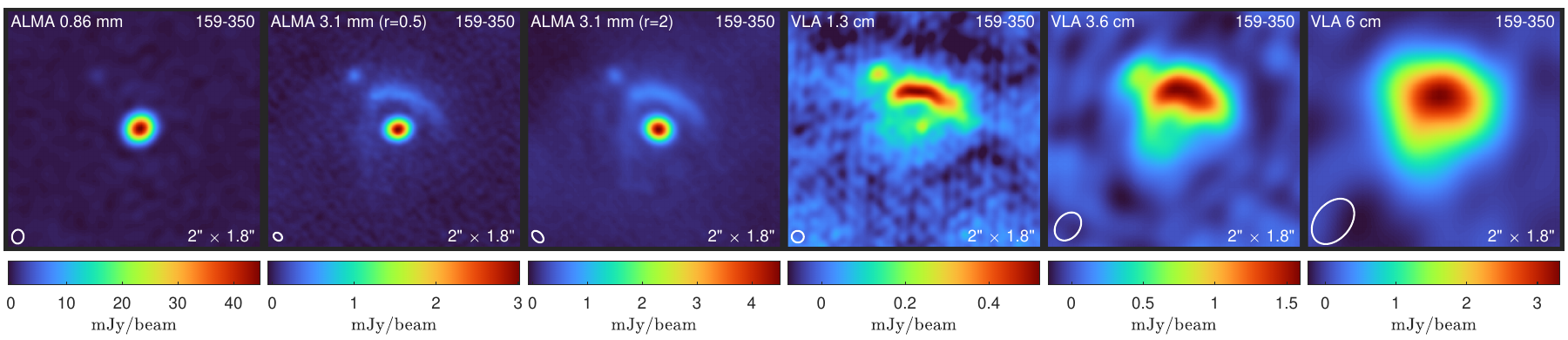}{1.0\textwidth}{}}\vspace{-10mm}
\gridline{\fig{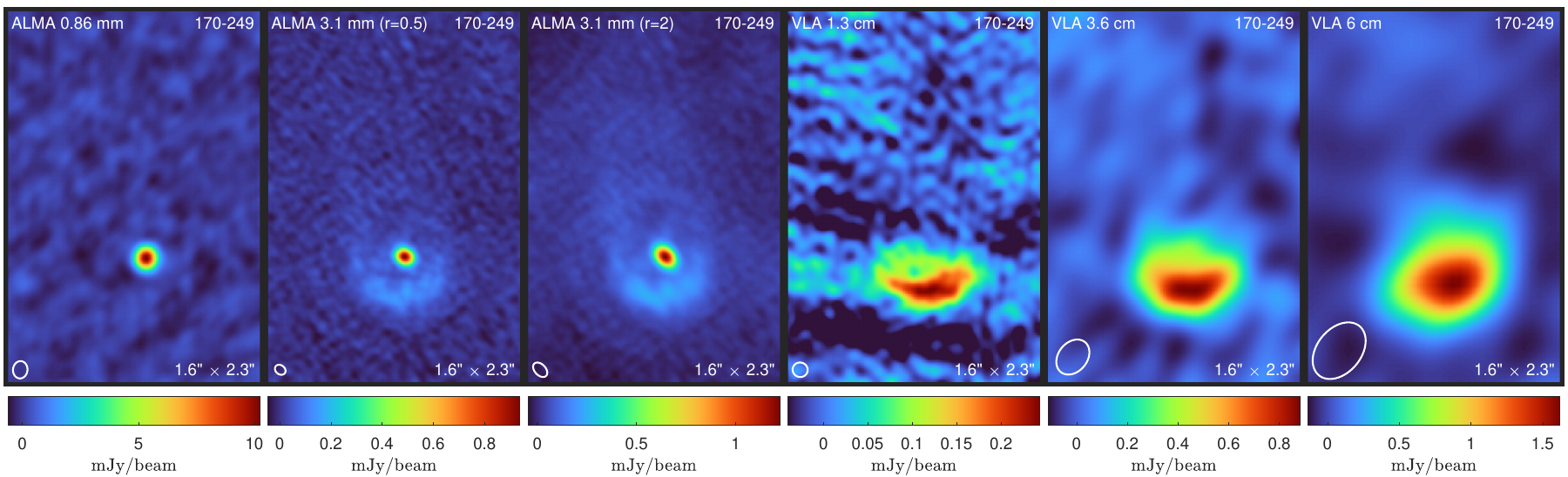}{1.0\textwidth}{}}\vspace{-10mm}
\gridline{\fig{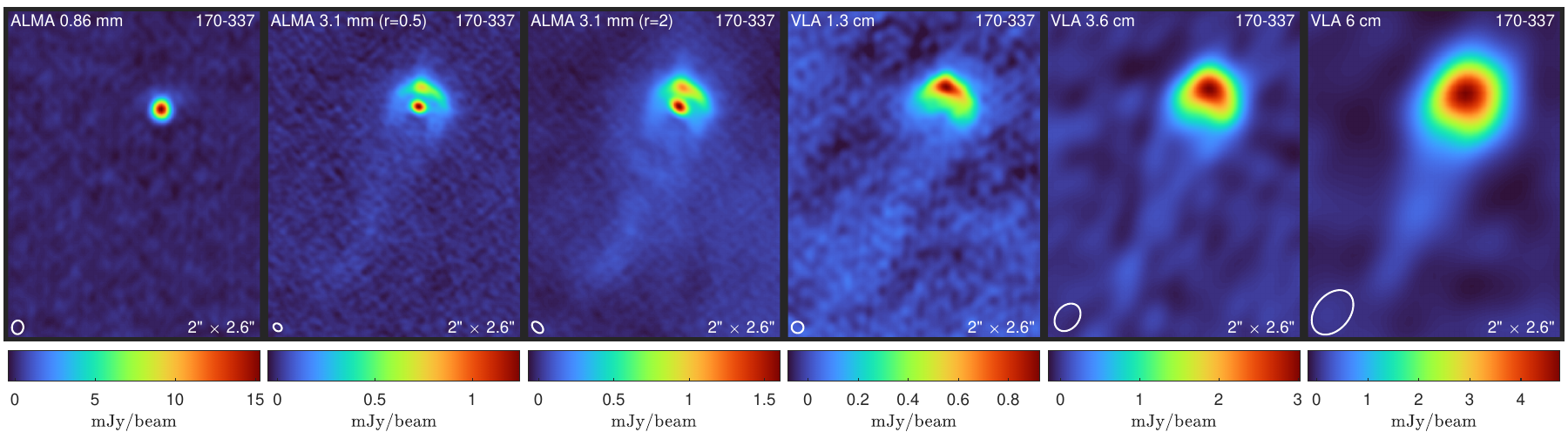}{1.0\textwidth}{}}\vspace{-10mm}
\gridline{\fig{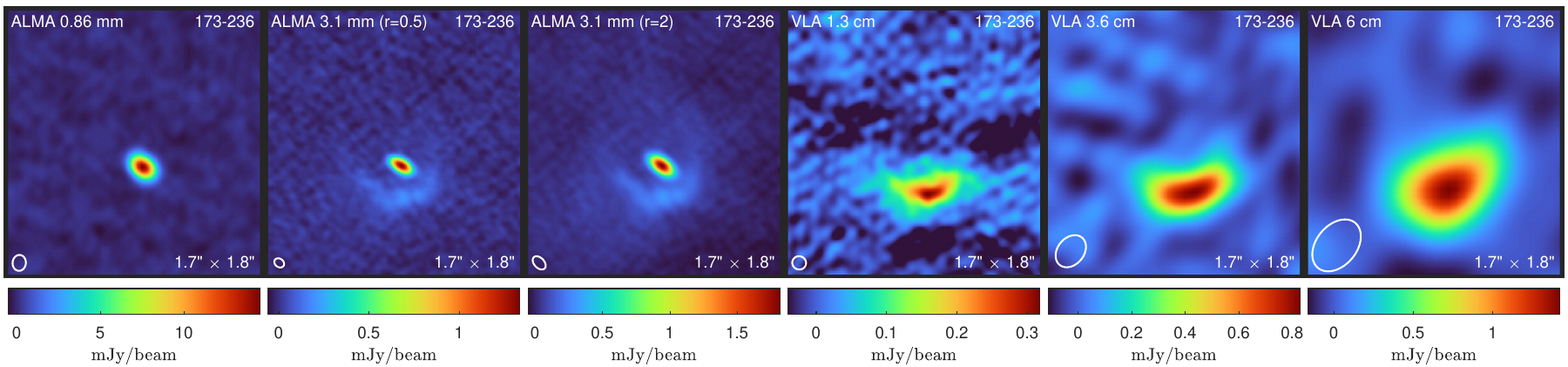}{1.0\textwidth}{}}\vspace{-10mm}
\gridline{\fig{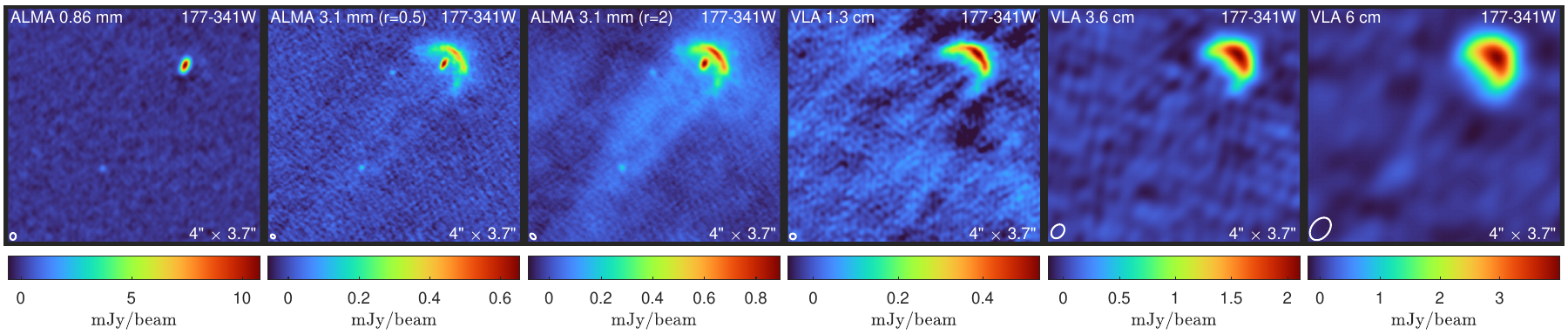}{1.0\textwidth}{}}
\caption{Same as Figure \ref{fig:proplyds1} but for proplyds 159-350, 170-249, 170-337, 173-236, and 177-341W. Note that the $K$ Band (1.3 cm) image of 159-350 detects the disk component in addition to the ionization front.}\label{fig:proplyds2}
\end{figure*}

\begin{figure*}
\gridline{\fig{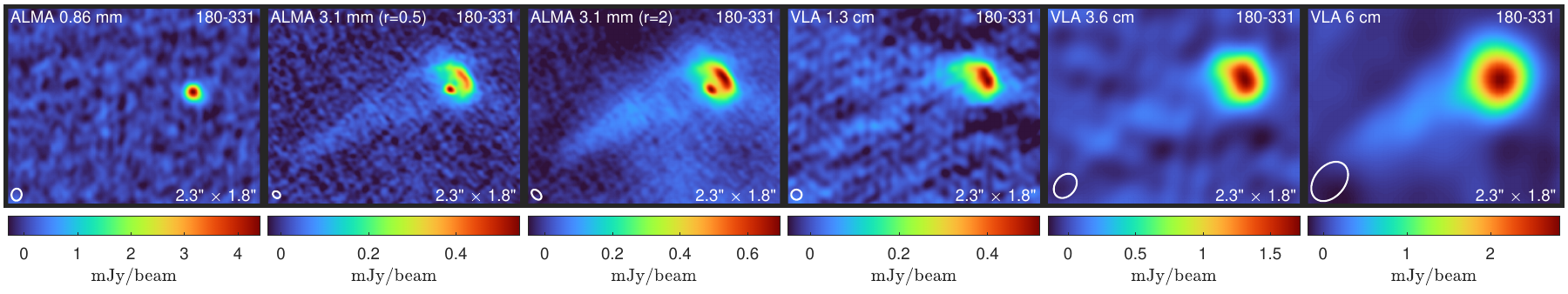}{1.0\textwidth}{}}\vspace{-10mm}
\gridline{\fig{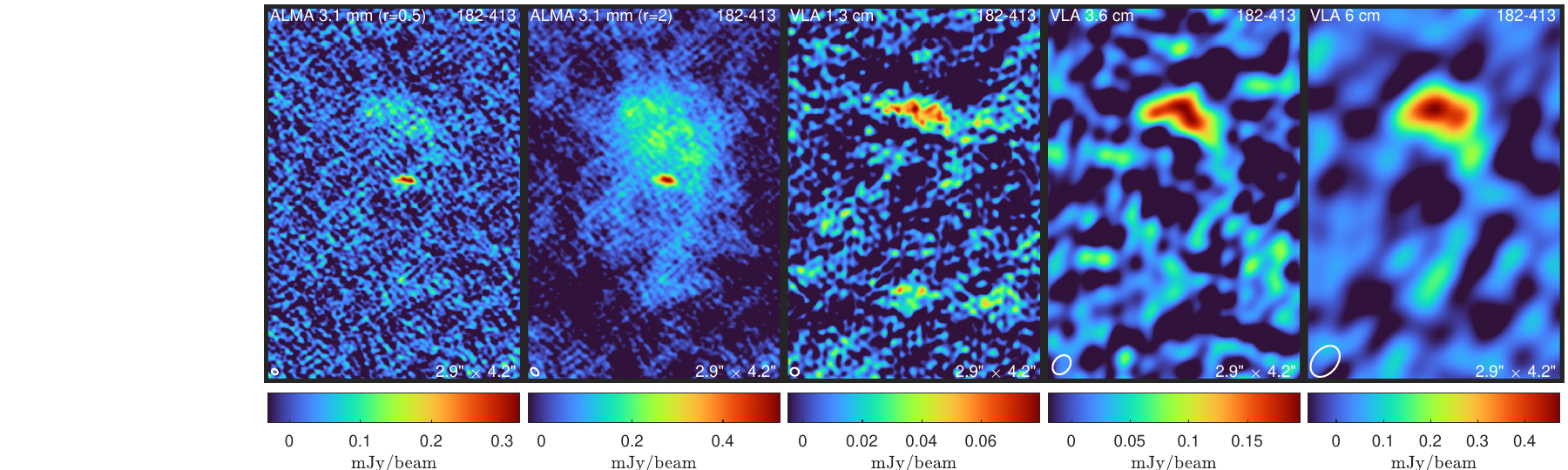}{1.0\textwidth}{}}
\caption{Same as Figures \ref{fig:proplyds1} and \ref{fig:proplyds2} but for proplyds 180-331 and 182-413. Proplyd 182-413 resides outside the bounds of the ALMA Band 7 mosaic.}\label{fig:proplyds3}
\end{figure*}

We identify over 200 sources in our final Band 3 mosaic. In this paper we focus only on the 12 proplyd systems that clearly exhibit both compact emission from a disk and free-free emission in a comet-like structure. An analysis of the full sample will be presented in future work. These 12 sources are well-known proplyds from HST images, and their locations are marked in Figure \ref{fig:field}. We refer to the proplyds by the six-digit coordinate-based naming scheme introduced by \citet{odell1994_HSTONC}, where the name XXX-YYY corresponds to a source in a $1\farcs5 \times 1\arcsec$ box whose northwest corner is at R.A. of 5:35:XX.X and decl. of -5:2Y:YY. In Figures \ref{fig:proplyds1}, \ref{fig:proplyds2}, and \ref{fig:proplyds3} we show the submillimeter through centimeter morphology of these proplyds.

To measure the disk component of each proplyd, we fit a 2D Gaussian to the central source using the CASA task imfit. Including a zero level offset in the fit yielded residuals with minimal structure aside from the expected free-free extended emission. The disk coordinates, deconvolved size, PA, and total flux density of each source are reported in Table \ref{tab:measured}. All 12 disks were successfully deconvolved in this manner. Additional properties of these systems, derived from the measurements, are presented in Table \ref{tab:derived}. 

We measure the Band 7 flux densities of these sources using imfit as well, and we report the results in Table \ref{tab:measured}. \citet{eisner2018_ALMAB7} measured Band 7 flux densities from the same data using the peak surface brightness rather than Gaussian fitting. To ensure a consistent measurement approach when comparing Band 7 and Band 3 (Section \ref{sec:specind}), we opt to use only the Band 7 values measured with imfit. These are typically higher than the fluxes reported by \citet{eisner2018_ALMAB7}, especially for the larger sources. We lack Band 7 measurements for one source, 182-413, because it resides outside the Band 7 mosaic.

\begin{figure*}
\epsscale{1.18}
\plotone{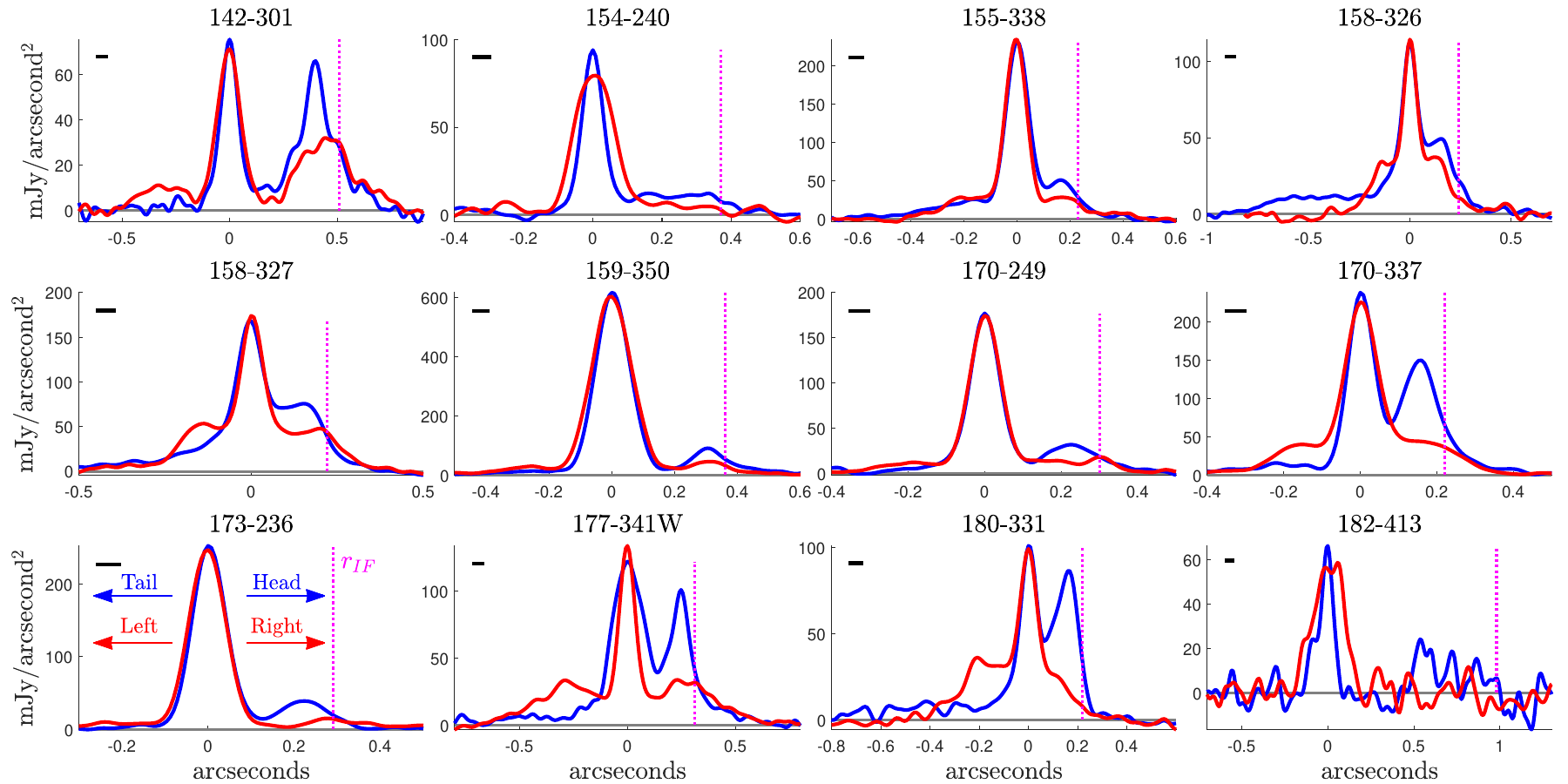}
\caption{Surface brightness cuts across the Band 3 (r=0.5) images of the proplyds. Each cut is the mean value of a 9-pixel (0\farcs072) wide strip centered on the disk. Blue lines are head-to-tail cuts with the origin at the disk and positive values toward the head. Red lines are horizontal cuts when the proplyd is rotated with the head facing up. The vertical dotted magenta line shows the location of the outer edge of the ionization front ($r_\text{IF}$). The black line in the upper left indicates the minor axis of the beam (0\farcs057).}
\label{fig:cuts}
\end{figure*}

We generate surface brightness cuts along and across the proplyds, as shown in Figure \ref{fig:cuts}. Each cut is from a 9 pixel (0\farcs072) wide strip centered on the disk. We made the cuts from the Band 3 (r=0.5) images after rotating them so the proplyd head faced up. Each head-to-tail cut (blue line) shows a prominent peak from the disk and a secondary peak at positive separations arising from the ionization front with brightness and width that varies considerably among the sources. For some of the sources (e.g., 155-38, 158-326, and 158-327) flux from the proplyd tail can also be seen in these cuts at negative values. The left-to-right cuts (red lines) show the disk and, in some sources, the sides of the ionization front.

We measure the size of the ionization front, which can be used to infer the photoevaporative mass-loss rate, as we will discuss is Section \ref{sec:Mdot}. Specifically, we measure $r_\text{IF}$, the distance from the disk center to the outer edge of free-free emission along the direction directly away from the proplyd tail. We do so using the surface brightness cuts, specifying  $r_\text{IF}$ where the blue curve shows an inflection point beyond its peak, identified by eye. We mark these locations with vertical dotted magenta lines in Figure \ref{fig:cuts} and report the values in Table \ref{tab:measured}.

We compute the projected distance on the sky ($d_\perp$) of the disk from $\theta^1$Ori C, the primary source of ionizing photons in the ONC. This will underestimate the true separation by, on average, 30\% \citep{johnstone1998_photoevap}.

Using spectral types and stellar luminosity measurements from \citet{fang2021_ONCHR}, we derive stellar masses for four of these 12 systems (158-326, 159-350, 170-249, and 173-236). The stellar types for the remaining sources are highly uncertain due to spectral veiling. We adopt stellar masses for four additional sources (142-301, 170-337, 177-341W, and 180-331) measured dynamically using ALMA observations of the gas disks \citep{boyden2023_ONCgas}. We report the stellar masses in Table \ref{tab:measured}.

To search for trends among the properties of these proplyds, we employ the Spearman rank correlation coefficient ($\rho$). This metric tests whether two sets of measurements are correlated but is agnostic about the functional form of the correlation. We list the correlation coefficients between the measured proplyd properties in Table \ref{tab:spearman}. For variables with measured uncertainties, we propagate the uncertainty to $\rho$ using a Monte Carlo sampling of the variables, and the listed values are the mean and standard deviation of the resulting distribution. We deem a relation to be a significant correlation if $\rho$ $>$ 0.5 at the 3$\sigma$ level, and we highlight those results in bold.    

\begin{deluxetable*}{lccccccccc}
\tablecaption{Measured Proplyd Properties \label{tab:measured}}
\tablewidth{0pt}
\tablehead{\colhead{Name} & \colhead{$M_\star$} & \colhead{R.A.} & \colhead{Decl.} & \colhead{Decon. $\theta_\text{maj}$} & \colhead{Decon. $\theta_\text{min}$} & \colhead{PA} & \colhead{B3 $F_\nu$} & \colhead{B7 $F_\nu$} & \colhead{$r_\text{IF}$} \\ \colhead{} & \colhead{($M_\odot$)} & \colhead{(J2000)} & \colhead{(J2000)} & \colhead{(\arcsec)} & \colhead{(\arcsec)} & \colhead{(deg)} & \colhead{(mJy)} & \colhead{(mJy)} & \colhead{(\arcsec)} \\ \colhead{(1)} & \colhead{(2)} & \colhead{(3)} & \colhead{(4)} & \colhead{(5)} & \colhead{(6)} & \colhead{(7)} & \colhead{(8)} & \colhead{(9)} & \colhead{(10)}}
\startdata
142-301 & 0.75 & 5 35 14.14 & -5 23 01.07 & 0.159 $\pm$ 0.011 & 0.035 $\pm$ 0.008 & 179 $\pm$ 2 & 1.12 $\pm$ 0.07 & 16.3 $\pm$ 0.6 & 0.51 \\ 
154-240 & \nodata & 5 35 15.39 & -5 22 39.85 & 0.108 $\pm$ 0.006 & 0.040 $\pm$ 0.004 & 84 $\pm$ 3 & 1.04 $\pm$ 0.04 & 10.8 $\pm$ 0.7 & 0.37 \\ 
155-338 & \nodata & 5 35 15.52 & -5 23 37.40 & 0.070 $\pm$ 0.004 & 0.065 $\pm$ 0.004 & 164 $\pm$ 62 & 2.41 $\pm$ 0.06 & 24.4 $\pm$ 0.2 & 0.23 \\ 
158-326 & 0.29 & 5 35 15.85 & -5 23 25.58 & 0.032 $\pm$ 0.010 & 0.025 $\pm$ 0.021 & 129 $\pm$ 79 & 0.55 $\pm$ 0.04 & 4.3 $\pm$ 0.2 & 0.24 \\ 
158-327 & \nodata & 5 35 15.80 & -5 23 26.57 & 0.044 $\pm$ 0.011 & 0.035 $\pm$ 0.016 & 15 $\pm$ 78 & 0.97 $\pm$ 0.08 & 10.4 $\pm$ 0.3 & 0.22 \\ 
159-350 & 0.73 & 5 35 15.95 & -5 23 50.04 & 0.132 $\pm$ 0.002 & 0.105 $\pm$ 0.002 & 109 $\pm$ 3 & 13.17 $\pm$ 0.15 & 131.3 $\pm$ 1.4 & 0.36 \\ 
170-249 & 0.10 & 5 35 16.97 & -5 22 48.52 & 0.066 $\pm$ 0.002 & 0.059 $\pm$ 0.002 & 68 $\pm$ 16 & 1.74 $\pm$ 0.03 & 14.3 $\pm$ 0.3 & 0.30 \\ 
170-337 & 0.75 & 5 35 16.99 & -5 23 37.06 & 0.070 $\pm$ 0.008 & 0.046 $\pm$ 0.008 & 85 $\pm$ 17 & 1.99 $\pm$ 0.13 & 21.7 $\pm$ 0.3 & 0.22 \\ 
173-236 & 0.93 & 5 35 17.35 & -5 22 35.74 & 0.106 $\pm$ 0.005 & 0.042 $\pm$ 0.003 & 60 $\pm$ 2 & 2.93 $\pm$ 0.10 & 26.1 $\pm$ 0.4 & 0.29 \\ 
177-341W & 0.75 & 5 35 17.69 & -5 23 40.97 & 0.152 $\pm$ 0.009 & 0.043 $\pm$ 0.007 & 152 $\pm$ 2 & 1.89 $\pm$ 0.11 & 21.2 $\pm$ 0.4 & 0.31 \\ 
180-331 & 0.40 & 5 35 18.05 & -5 23 30.82 & 0.067 $\pm$ 0.010 & 0.050 $\pm$ 0.010 & 69 $\pm$ 87 & 0.80 $\pm$ 0.07 & 6.0 $\pm$ 0.1 & 0.22 \\ 
182-413 & \nodata & 5 35 18.22 & -5 24 13.55 & 0.190 $\pm$ 0.025 & 0.061 $\pm$ 0.012 & 88 $\pm$ 4 & 1.33 $\pm$ 0.16 & \nodata & 0.98 \\ 
\enddata
\tablecomments{Column (1): proplyd name. Column (2): stellar mass, with 158-326, 159-350, 170-249, and 173-236 using spectral type and luminosity measurements from \citet{fang2021_ONCHR}, and 142-301, 170-337, 177-341W, and 180-331 using dynamical masses from \citet{boyden2023_ONCgas}. Columns (3) and (4): coordinates of the disk center from imfit. Columns (5) and (6): ceconvolved FWHM major and minor axes of the disk from imfit. Column (7): position angle of the deconvolved disk from imfit. Columns (8) and (9): flux density of the disk from imfit in Band 3 and Band 7. Uncertainties are the statistical errors returned by imfit. Column (10): distance from the disk to the outer edge of the ionization front measured in the direction opposite of the proplyd tail.}
\end{deluxetable*}

\begin{deluxetable*}{lcccccccccccC}
\tablecaption{Computed Proplyd Properties \label{tab:derived}}
\tablewidth{0pt}
\tabletypesize{\footnotesize}
\tablehead{\colhead{Name} & \colhead{$d_\perp$} &  \colhead{$r_g$} & \colhead{$r_d$} &  \colhead{$r_d/r_g$} & \colhead{$T_d$} & \colhead{$M_\text{d}$(20\,K)} & \colhead{$M_\text{d}(T_d)$} & \colhead{$\alpha_{7-3}$} & \colhead{$\dot{M}_1$} & \colhead{$\dot{M}_2$} & \colhead{$t_\textrm{min}$} & \colhead{$t_\textrm{max}$} \\ \colhead{} & \colhead{(\arcsec)} & \colhead{(au)} & \colhead{(au)} & \colhead{} & \colhead{(K)} & \colhead{($M_\oplus$)} & \colhead{($M_\oplus$)} & \colhead{}  & \colhead{($10^{-7}$\,$M_\odot$\,yr$^{-1}$)} & \colhead{($10^{-7}$\,$M_\odot$\,yr$^{-1}$)} & \colhead{(kyr)} & \colhead{(kyr)} \\ \colhead{(1)} & \colhead{(2)} & \colhead{(3)} & \colhead{(4)} & \colhead{(5)} & \colhead{(6)} & \colhead{(7)} & \colhead{(8)} & \colhead{(9)} & \colhead{(10)} & \colhead{(11)} & \colhead{(12)} & \colhead{(13)}}
\startdata
142-301 & 41.0 & 73.9 & 31.8 $\pm$ 2.2 & 0.43 & 68 & 55.4 $\pm$ 3.4 & 15.1 $\pm$ 0.9 & 2.09 $\pm$ 0.12 & 1.6 & 14.4 & 3.1 & 104 \\ 
154-240 & 46.0 & \nodata & 21.6 $\pm$ 1.1 & \nodata & 64 & 51.6 $\pm$ 2.2 & 14.9 $\pm$ 0.6 & 1.83 $\pm$ 0.13 & 0.9 & 3.4 & 13.3 & 176 \\ 
155-338 & 20.2 & \nodata & 14.1 $\pm$ 0.8 & \nodata & 96 & 119.6 $\pm$ 3.1 & 22.6 $\pm$ 0.6 & 1.81 $\pm$ 0.11 & 1.0 & 3.2 & 21.2 & 366 \\ 
158-326 & 9.6 & 28.6 & 6.4 $\pm$ 2.1 & 0.23 & 140 & 27.4 $\pm$ 2.1 & 3.5 $\pm$ 0.3 & 1.61 $\pm$ 0.13 & 2.2 & 3.3 & 3.2 & 37 \\ 
158-327 & 10.6 & \nodata & 8.8 $\pm$ 2.3 & \nodata & 133 & 48.2 $\pm$ 3.8 & 6.6 $\pm$ 0.5 & 1.86 $\pm$ 0.13 & 1.7 & 3.8 & 5.1 & 83 \\ 
159-350 & 28.2 & 72.0 & 26.4 $\pm$ 0.4 & 0.37 & 82 & 654.0 $\pm$ 7.7 & 146.6 $\pm$ 1.7 & 1.80 $\pm$ 0.11 & 1.4 & 9.6 & 45.8 & 1427 \\ 
170-249 & 35.2 & 9.9 & 13.2 $\pm$ 0.5 & 1.33 & 73 & 86.4 $\pm$ 1.5 & 21.7 $\pm$ 0.4 & 1.65 $\pm$ 0.11 & 0.8 & 3.5 & 18.5 & 310 \\ 
170-337 & 16.1 & 73.9 & 14.0 $\pm$ 1.7 & 0.19 & 108 & 99.0 $\pm$ 6.6 & 16.7 $\pm$ 1.1 & 1.87 $\pm$ 0.12 & 1.1 & 5.4 & 9.2 & 259 \\ 
173-236 & 49.0 & 91.7 & 21.2 $\pm$ 0.9 & 0.23 & 62 & 145.4 $\pm$ 4.9 & 43.4 $\pm$ 1.5 & 1.71 $\pm$ 0.11 & 0.6 & 4.1 & 31.9 & 763 \\ 
177-341W & 25.7 & 73.9 & 30.5 $\pm$ 1.9 & 0.41 & 85 & 93.6 $\pm$ 5.5 & 20.0 $\pm$ 1.2 & 1.89 $\pm$ 0.12 & 1.2 & 8.0 & 7.5 & 233 \\ 
180-331 & 25.0 & 39.4 & 13.4 $\pm$ 2.1 & 0.34 & 87 & 39.6 $\pm$ 3.4 & 8.3 $\pm$ 0.7 & 1.58 $\pm$ 0.13 & 0.7 & 4.2 & 5.9 & 160 \\ 
182-413 & 57.0 & \nodata & 37.9 $\pm$ 5.1 & \nodata & 57 & 66.3 $\pm$ 8.0 & 21.4 $\pm$ 2.6 & \nodata & 3.1 & 18.4 & 3.5 & 65 \\ 
\enddata
\tablecomments{Column (1): proplyd name. Column (2): projected separation between the disk center and $\theta^1$Ori C. Column (3): gravitational radius computed from the stellar mass assuming $c_s$ = 3 km s$^{-1}$. Column (4): disk radius from the deconvolved size of the major axis ($\theta_\text{maj}$) assuming a distance of 400 pc. Column (5): ratio of the disk radius to the gravitational radius. Column (6): disk dust temperature due to external heating by $\theta^1$ Ori C from Equation \ref{eq:Tdext}. Column (7): disk dust mass computed from the Band 3 flux density using Equation \ref{eq:flux-mass} and assuming a dust temperature of 20 K. Column (8): same as Column 7 but using a dust temperature ($T_d$) set by external heating. Uncertainties on both dust masses are propagated from the flux density uncertainties. Column (9): power-law spectral index of the disk measured from Band 7 to Band 3. Uncertainties include a 10\% calibration uncertainty on the Band 7 and Band 3 flux densities. Column (10): photoevaporative mass-loss rate computed from Equation \ref{eq:Mdot1}. Column (11): photoevaporative mass-loss rate computed from Equation \ref{eq:Mdot2}. Column (12) minimum estimate of the photoevaporation timescale computed as 100$M_\text{d}(T_d)/\dot M_\text{2}$. Column (13): maximum estimate of the  photoevaporation timescale computed as 100$M_\text{d}(20 \,K)/\dot M_\text{1}$.}
\end{deluxetable*}

\begin{deluxetable}{lc}
\tablecaption{Spearman Correlations \label{tab:spearman}}
\tablewidth{0pt}
\tablehead{\colhead{Relation} & \colhead{$\rho$}}
\startdata
$r_d$(B3) versus $r_d$(B7) & \textbf{0.933 $\pm$ 0.036} \\ 
$r_d$ versus $r_g$ & 0.615 $\pm$ 0.078 \\ 
$r_d$ versus $F_\nu$(B3) & 0.378 $\pm$ 0.081 \\ 
$r_d$ versus $r_\textrm{IF}$ & \textbf{0.795 $\pm$ 0.041} \\ 
$F_\nu$(B3) versus $r_\textrm{IF}$ & 0.131 $\pm$ 0.044 \\ 
$\alpha_{7-3}$ versus $r_d$ & 0.441 $\pm$ 0.204 \\ 
$\alpha_{7-3}$ versus $F_\nu$(B3) & 0.181 $\pm$ 0.214 \\ 
$r_d$ versus $d_\perp$ & \textbf{0.741 $\pm$ 0.036} \\ 
$F_\nu$(B3) versus $d_\perp$ & 0.290 $\pm$ 0.039 \\ 
$r_\textrm{IF}$ versus $d_\perp$ & \textbf{0.775} \\ 
\enddata
\tablecomments{Relations with correlation coefficient $\rho$ $>$ 0.5 at 3$\sigma$ significance are listed in bold.}
\end{deluxetable}

\section{Results and Discussion}
\label{sec:results}

\subsection{Wavelength-dependent Morphology}

The 12 proplyds in our sample show a consistent pattern of morphology versus wavelength. At Band 7, only emission from the central disk is seen. At Band 3, both compact emission from the disk and extended emission from the ionization front are seen. Free-free emission from the proplyd tail is also seen in some sources, and it is more clear in the r=2.0 images. At centimeter wavelengths, the ionization front is clearly detected, but the disk is not seen (with the exception of 159-350). The pattern is due to the spectral dependence of thermal dust emission compared to free-free emission. The flux density of a dust disk decreases with increasing wavelength, following $F_\nu \propto \lambda^{-\alpha}$ with spectral index $\alpha$ = 1.5--3, making it difficult to detect at centimeter wavelengths. The spectral dependence of (optically thin) free-free emission is shallow, following $F_\nu \propto \lambda^{0.1}$, so its brightness dominates over dust at longer wavelengths. This behavior has been employed in previous studies to spectrally decompose dust from free-free emission for unresolved sources with no obvious propyld morphology \citep[e.g.,][]{mann2014_ALMA7ONC,sheehan2016_ONCVLA,eisner2018_ALMAB7}. With deep, high-resolution images of proplyds, as presented here, we can spatially isolate the dust disk from the surrounding free-free emission.

\subsection{Additional Observed Features}

\begin{figure}
\epsscale{1.17}
\plotone{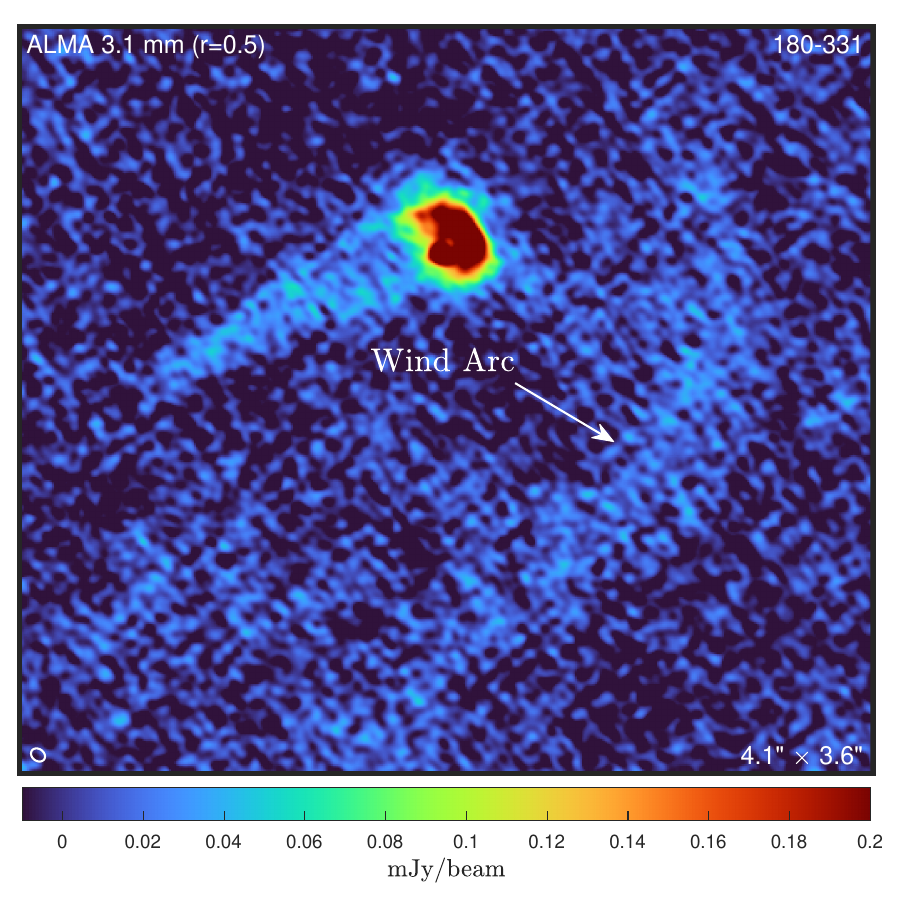}
\caption{ALMA Band 3 image of the proplyd 180-331 showing the wind arc, which was also seen in HST images of this source. This is the only proplyd in our sample with a wind arc detected. The color scale is truncated, saturating the disk and proplyd head.}
\label{fig:arc}
\end{figure}

HST images found arcs of emission beyond the ionization front of some proplyds \citep{bally1998_ONCHST}. These are bow shocks from the interaction of the outflowing material with the stellar wind of $\theta^1$ Ori C. We searched the region around these 12 proplyds in the Band 3 image for such arcs, and we located one around 180-331 (Figure \ref{fig:arc}). The arc is faint but detected as a coherent structure over the background noise, and it is coincident with the arc seen with HST.

The ALMA Band 3 image of 170-337 exhibits a ``knot" of emission in the ionization front directly north of the disk. This marks the location where a jet from the disk breaks through the ionization front. The jet is known from prior HST observations \citep{odell1997_Orionhighv,bally1998_ONCHST,bally2000_ONC}. Faint emission is seen extending from the knot beyond the ionization front, likely tracing free-free emission from jet material that has been photoionized.

Two compact sources are observed near 177-341W in the ALMA Band 3 observations. One, east of the disk, was discovered in HST images by \citet{bally1998_ONCHST} and designated 177-341b. This source is not seen in Band 7 but is perhaps faintly visible in the VLA $K$ and $X$ band images. \citet{bally2000_ONC} suggests this source may be a knot of emission caused by a jet or outflow from the disk in this proplyd. The second compact source is southeast of the disk near the end of the proplyd tail. This source shows opposite panchromatic behavior---it is clearly seen in ALMA Band 7 but not apparent at centimeter wavelengths, suggesting the emission is primarily from dust. 159-350 also has a companion source located northeast of the ionization front. It is seen in all wavelengths presented here, as well as in HST images \citep{bally1998_ONCHST}.  

\subsection{Disk Sizes}
\label{sec:DiskSizes}

The ability of these observations to isolate the dust disks from the surrounding free-free emission allows us to study their sizes and masses, which are fundamental in determining the nature of the planetary system that can form.  

We compute the disk radii as one-half of the FWHM of the deconvolved major axis, assuming a distance of 400 pc, and we report the results in Table \ref{tab:derived}. Disk radii range from 6.4 to 38 au, which are comparable to, or smaller than, the size of the solar system ($\sim$30 au), suggesting these disks will form compact planetary systems. The radii are consistent with typical dust disk sizes seen in the broader ONC population with ALMA \citep{eisner2018_ALMAB7,otter2021_OMC}, providing further evidence that external photoevaporation---which is clearly impacting these proplyd systems---sets the disk sizes in the ONC more generally.

\begin{figure}
\epsscale{1.17}
\plotone{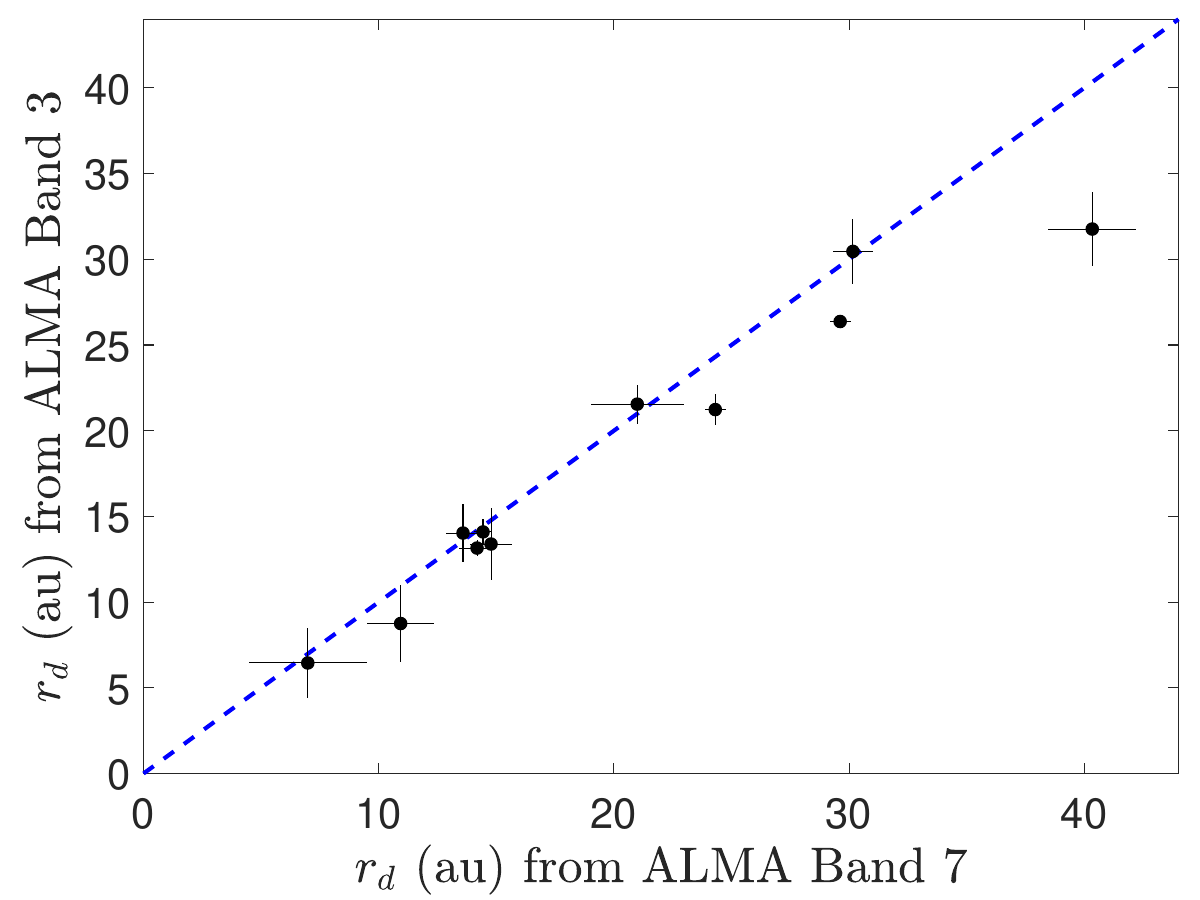}
\caption{Size of the proplyd disks measured from ALMA Band 3 and Band 7 images. The sizes agree for most sources although three are slightly smaller at Band 3. Proplyd 182-413 was not covered by the Band 7 observations and thus is excluded from this plot.}
\label{fig:Rd3vs7}
\end{figure}

We compare the sizes of the proplyd disks measured at Band 3 and Band 7 (Figure \ref{fig:Rd3vs7}). We find very good agreement for most of the disks ($\rho$ = 0.933 $\pm$ 0.036), although three are slightly smaller at Band 3. A smaller disk size at longer wavelengths is expected if larger grains (which emit more efficiently at longer wavelengths) are concentrated to smaller disk radii than smaller grains, as predicted by inward aerodynamic drift \citep{weidenschilling1977_drift}. The absence of a size difference between wavelengths could indicate that drift is halted by, i.e., pressure bumps in the disk \citep{pinilla2012_trapping}, that grains at all radii are sufficiently large to emit efficiently out to 3.1 mm, or that the disk is optically thick. High optical depth is likely for these systems (Section \ref{sec:specind}). Disks having similar sizes at Band 3 and Band 7 are not unique to the ONC; \citet{tazzari2021_3mmsizes} found this to be the case for many disks in the Lupus star-forming region. 

\begin{figure}
\epsscale{1.17}
\plotone{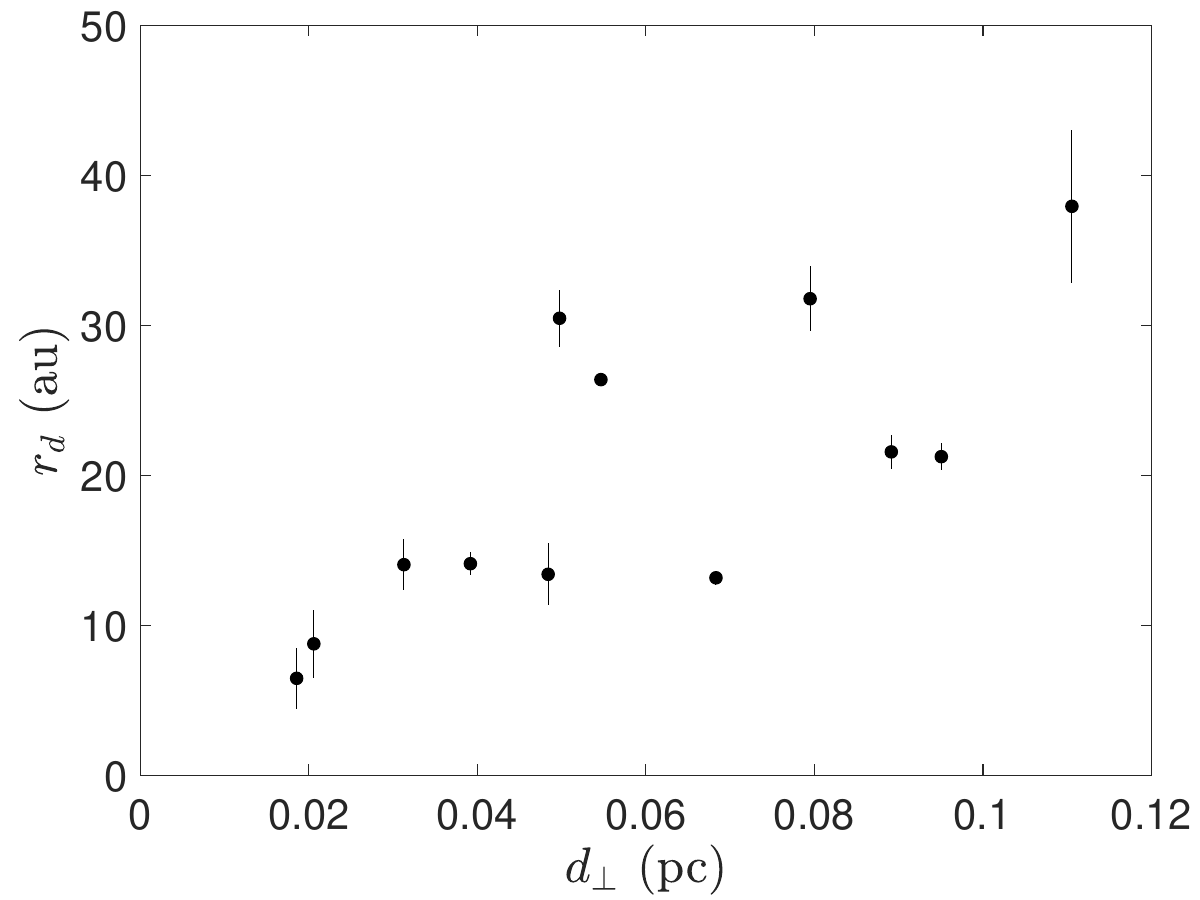}
\caption{Size of the proplyd disks versus their projected separation from $\theta^1$ Ori C.}
\label{fig:RdDperp}
\end{figure}

\begin{figure}
\epsscale{1.17}
\plotone{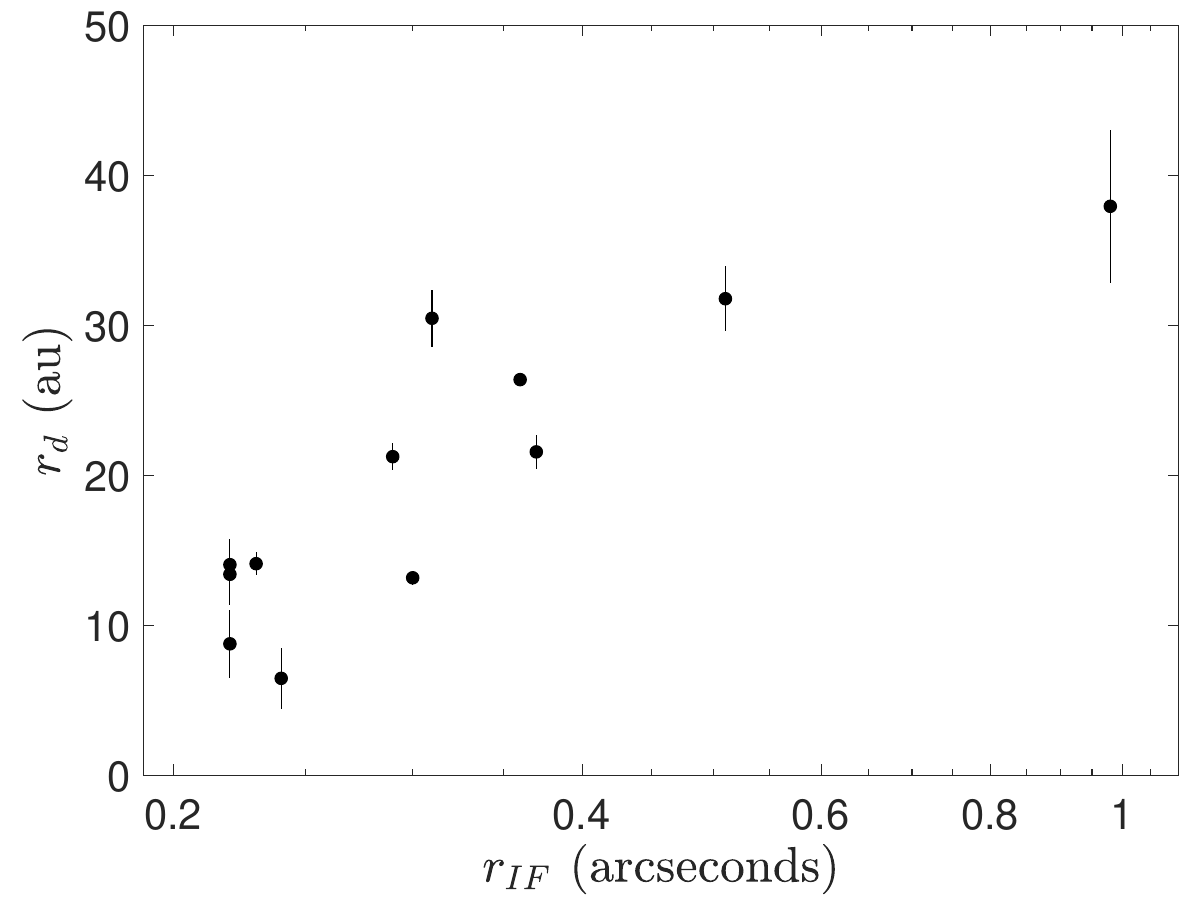}
\caption{Size of the proplyd disks versus size of the proplyd ionization fronts.}
\label{fig:RdRIF}
\end{figure}

We identify a correlation between $r_d$ and $d_\perp$ with $\rho$ = 0.741 $\pm$ 0.036 (Figure \ref{fig:RdDperp}). This differs from the results of the larger ONC samples presented by \citet{eisner2018_ALMAB7} and \citet{otter2021_OMC}, who found no significant trend. \citet{boyden2020_ONCgas} did, however, find a trend in gas disk radii with $d_\perp$. There is also a clear trend between $r_d$ and $r_\textrm{IF}$ with $\rho$ = 0.795 $\pm$ 0.041 (Figure \ref{fig:RdRIF}), which was previously seen in HST images of proplyds \citep{vicente2005_ONCdisks}.

In theory, external photoevaporation removes material from the disk beyond the ``gravitational radius" ($r_g$). This the location where the sound speed of the gas ($c_s$) is equal to the escape speed from the star's gravity, i.e., $r_g = GM_\star / c_s^2$. We compute $r_g$ for the eight systems with measured stellar masses (see Table \ref{tab:measured}) using $c_s$ = 3 km s$^{-1}$ as appropriate for an FUV-driven outflow \citep{johnstone1998_photoevap,winter2022_photoevapreview} and report the results in Table \ref{tab:derived}. We find $r_d/r_g$ ratios ranging from 1.33 to 0.19. The one system with $r_d$ $>$ $r_g$ (170-249) is undergoing photoevaporation in the ``supercritical" regime \citep{adams2004_photoevap} and may be at an earlier stage of the photoevaporation process. Systems with $r_d$ $<$ $r_g$ (but $r_d/r_g$ $\gtrsim$ 0.1--0.2) are consistent with results of detailed photoevaporation models in the ``subcritical" regime \citep{adams2004_photoevap}. We note that our observations only probe the radii of the dust disks, which can be smaller than those of the gas disks if large grains have drifted inward. Gas disks larger than dust disks is regularly observed in both low-mass star-forming regions \citep{ansdell2018_gasradii} and the ONC \citep{boyden2020_ONCgas}. 

\subsection{Disk Spectral Indices and Optical Depths}
\label{sec:specind}

\begin{figure}
\epsscale{1.17}
\plotone{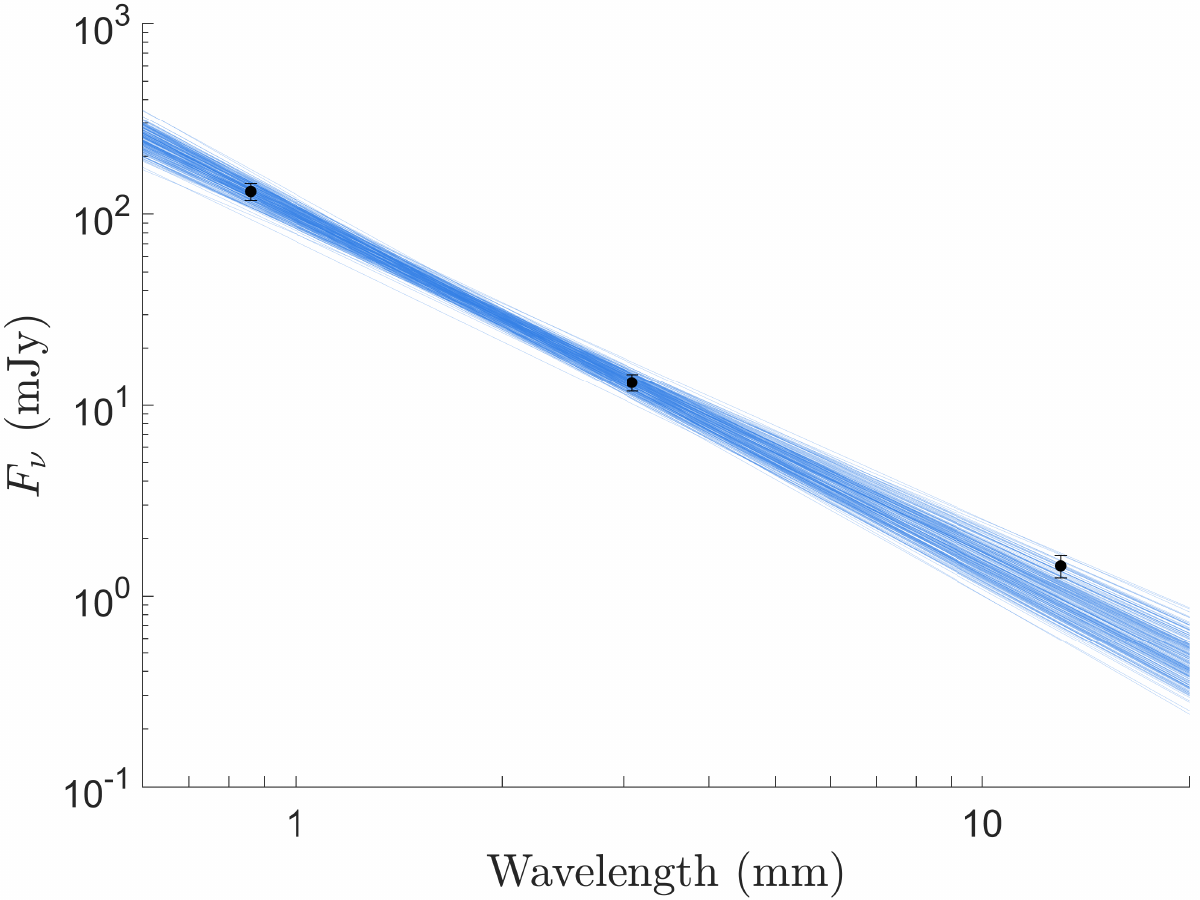}
\caption{SED of the 159-350 disk, the only system in our sample with the disk component detected at VLA $K$ band. Black points are flux density measurements from ALMA Band 7, ALMA Band 3, and VLA $K$ Band. Error bars represent the total (measured plus calibration) uncertainty. Blue lines are 200 power-law fits to the two ALMA points while varying them within their uncertainties. The $K$ band flux density is consistent with these fits (and thus with arising from dust emission) although it is on the brighter end of the expected range.}
\label{fig:SED}
\end{figure}

The spectral dependence of disk flux density in the (sub)millimeter typically follows a power law $F_\nu \propto \lambda^{-\alpha}$. If the emission is optically thin and in the Rayleigh-Jeans regime, measuring the spectral index ($\alpha$) can reveal the index of the dust opacity spectrum ($\beta = \alpha-2$). The spectral index can thus inform us of the dust grain sizes, with shallower spectral indices indicating growth to larger grains. If the emission is optically thick, scattering can be significant, and the spectral index is shallower (closer to 2) and depends on the dust albedo \citep{liu2019_PPscat,zhu2019_PPscattering}.

We measure the spectral indices of the disks from Band 7 to Band 3 for the eleven sources observed in both bands. Values are reported in Table \ref{tab:derived}. The uncertainties on $\alpha$ are propagated from statistical errors on the flux values returned from imfit plus a 10\% error due to calibration uncertainty. We find $\alpha$ values from 1.6 to 2.1, strongly suggesting the emission is optically thick. If scattering is ignored, optically thick emission follows the Planck function, so $\alpha$ $<$ 2 requires cold dust in order to deviate from the Rayleigh-Jeans regime. For instance, the smallest spectral index in our sample, $\alpha$ = 1.61, would require a dust temperature of 14 K, which is colder than expected for disks in isolated environments, much less in the ONC. It is much more likely that $\alpha$ $<$ 2 is due to the scattering of radiation by dust with albedo that decreases with increasing wavelength \citep{liu2019_PPscat,zhu2019_PPscattering}.

In principle, this behavior constrains the grain properties. \citet{zhu2019_PPscattering} show that $\alpha$ $<$ 2 arises when the maximum grain size is around a few hundred microns. This is consistent with model predictions for dust at 10s of astronomical units in a 1 Myr old disk, as shown, for example, in Figure 5 of \citet{rosotti2019_radiievolution}. Models that include external photoevaporation, however, predict a reduction in the maximum grain size \citep{sellek2020_dustphotoevap}. A detailed investigation into the grain properties of these disks could be carried out in a future paper.

\citet{otter2021_OMC} found a large population of ONC sources with $\alpha \simeq 2$, consistent with our results. \citet{tazzari2021_3mmLupus} measured $\alpha$ for disks in the Taurus, Lupus, and Ophiuchus low-mass star-forming regions, finding values from 1.7 to 2.9. The proplyd disk indices reported here fall into the lower end of this range. The high optical depth of ONC disks is likely a result of their small sizes. \citet{tazzari2021_3mmsizes} found a trend of increasing $\alpha$ with disk size for systems in Lupus. We find no clear correlation between $\alpha$ and $r_d$ within our sample (Table \ref{tab:spearman}), although the range of disk sizes over which to search for such a trend is limited. Studies of large and well-resolved disks have found $\alpha$ to be $<$ 2 in the inner (presumably more optically thick) regions \citep{tsukagoshi2016_TWhya,huang2018_TWhya,dent2019_submmpol,ribas2023_MPMus}. The proplyd disks may be similar, only with the outer parts removed by external photoevaporation.  

For one source, 159-350, flux is clearly detected in the VLA $K$ Band (1.3 cm) at the location of the disk. The flux density is 1.44 $\pm$ 0.19 mJy, as found by imfit. We examine its spectral energy distribution (SED; Figure \ref{fig:SED}) to assess whether this emission arises from dust, free-free, or a nonthermal source of emission. Blue lines show a set of 200 power-law fits to the Band 7 and Band 3 flux densities, varied independently within their uncertainties. The $K$ band measurement is consistent with (although at the higher end of) these fits extrapolated to 1.3 cm. Thus, we conclude this emission comes primarily from dust, and that this disk likely remains optically thick to centimeter wavelengths. This disk is, by far, the brightest in our sample (4.5 times brighter than the second-brightest disk in ALMA Band 3), which explains why it is the only one detected with the VLA. 

For some of the other disks, extrapolating the measured Band 3 fluxes to $K$ Band using the measured Band 7--3 spectral index predicts faint, but significant, detections. However, no clear disk components are observed. This suggests a steepening of the spectral index between 3.1 mm and 1.3 cm, likely due to the disks becoming optically thin or a change in the spectral dependence of the dust albedo. Future observations at wavelengths longer than 3.1 mm but sensitive enough to detect the dust disks would be useful to test this. Such measurements could come from deeper VLA observations, upcoming observations with ALMA Band 1, or future observations with the ngVLA.     

\subsection{Disk Masses and Temperatures}
\label{sec:DiskMasses}

We compute the disk dust masses with the standard equation
\begin{equation}
\label{eq:flux-mass}
M_\text{d} = \frac{F_\nu d^2}{\kappa B_\nu\left(T_\text{d}\right)}
\end{equation}
\citep{hildebrand1983_dustmass,beckwith1990_disksurvey}, where $F_\nu$ is the measured Band 3 flux density and $d$ is the distance from Earth (400 pc). $\kappa$ is the dust opacity, which we set to 1 cm$^2$ g$^{-1}$ based on the value of 2.3 cm$^2$ g$^{-1}$ used by \citet{andrews2013_MdMstar} at 1.3 mm and the commonly assumed spectral dependence $\kappa \propto \lambda^{-1}$. $B_\nu\left(T_\text{d}\right)$ is the Planck function at the average dust temperature, for which we use two different choices. First, we use 20 K, which is the typically assumed dust temperature for disks in low-mass star-forming regions. Second, we use the temperature expected due to external heating by $\theta^1$ Ori C \citep{haworth2021_warmdust}. From balancing disk heating and cooling, this temperature is
\begin{equation}
\label{eq:Tdext}
    T_d = T_\star \left(\frac{R_\star}{d_\perp}\right)^{1/2},
\end{equation}
where $T_\star$ = 39,000 K and $R_\star$ = 10.6 $R_\odot$ are the temperature and radius of $\theta^1$ Ori C, respectively. We find $T_d$ $>$ 20 K in all cases, suggesting external heating dominates over internal heating when setting the disk temperatures. We report $T_d$ and both mass calculations in Table \ref{tab:derived}. The uncertainties we quote on the dust masses are simply the propagation of the uncertainty on the flux densities.

Assuming 20 K, disk dust masses range from 27 to 654 $M_\oplus$, with all but one of the disks having $M_d$ $>$ 33 $M_\oplus$, the mass of dust in models of the Minimum Mass Solar Nebula \citep{weidenschilling1977_MMSN}. This suggests that these disks are capable of forming planetary systems like the solar system. With $T_d$ set by external heating, however, dust masses are lower (3.5--147 $M_\oplus$), and only two disks have $M_d$ $>$ 33 $M_\oplus$. We find no correlation between disk mass (actually disk flux) and $d_\perp$ (Table \ref{tab:spearman}), in agreement with the findings by \citet{otter2021_OMC} for a larger sample of ONC sources.

\begin{figure}
\epsscale{1.17}
\plotone{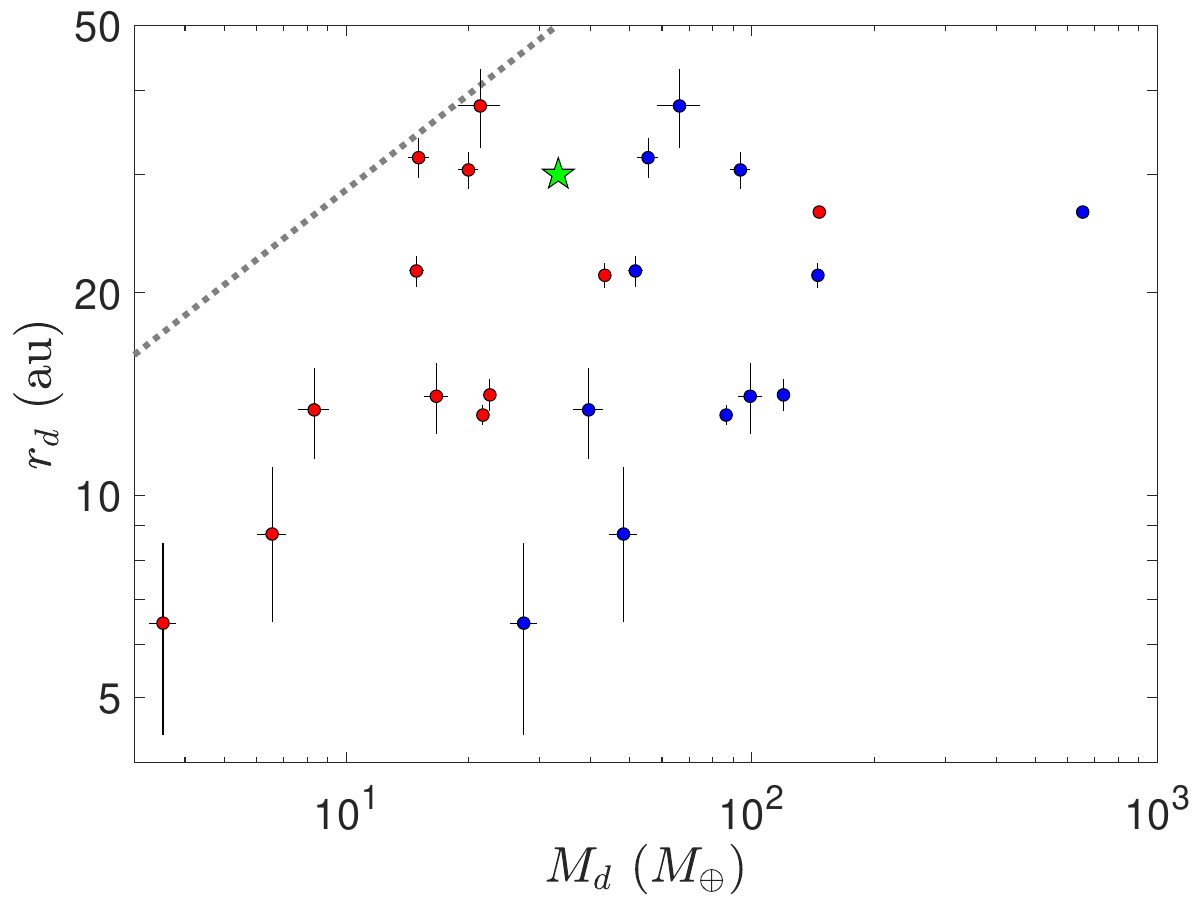}
\caption{Disk radii and dust masses of the twelve proplyds in this sample. Masses computed assuming 20 K dust are in blue, whereas masses computed using a dust temperature set by external radiation ($T_d$) are in red. The green star is the Minimum Mass Solar Nebula. The gray dotted line is the disk size--mass relation found by \citet{tripathi2017_sizemass} in low-mass star-forming regions.}
\label{fig:RdMd}
\end{figure}

In Figure \ref{fig:RdMd} we plot disk radii versus dust mass, and we see no significant trend between these properties. This is in contrast to low-mass star-forming regions where a trend does exist between disk size and mass \citep{tripathi2017_sizemass,andrews2018_sizemass}, which we show as a gray dotted line in Figure \ref{fig:RdMd}. We see that the proplyd disks are smaller than predicted by the trend established for disks in low-mass star-forming regions. No size--mass (or size--flux) trend was seen in prior ALMA studies of larger samples of ONC disks \citep{eisner2018_ALMAB7,otter2021_OMC}. We also note that the proplyd disks are brighter than most other disks in the ONC, yielding a biased view of the size--mass parameter space relative to these prior studies.

Equation \ref{eq:flux-mass} assumes the dust emission is entirely optical thin. As discussed in Section \ref{sec:specind}, the measured spectral indices suggest high optical depth, leading the masses to be underestimated \citep{ballering2019_taurusSEDs,xin2023_lupusSEDs}. The results are still useful as a lower limit on the dust masses and for comparison with masses of other disks calculated similarly. The effect on disk dust masses due to both temperature and optical depth effects can be accounted for with radiative transfer models \citep[e.g.,][]{ballering2019_taurusSEDs}, but such models are beyond the scope of this work.

\subsection{Mass-loss Rates}
\label{sec:Mdot}

Measuring the photoevaporative mass-loss rate ($\dot M$) is critical to understanding the effect of the cluster environment on the planet-forming potential in disks. The FUV-driven wind from the disk surface is roughly symmetric \citep{winter2022_photoevapreview}, so we compute the mass flow through a spherical surface at the location of the ionization front: $\dot M = 4 \pi r_\text{IF}^2 \rho v$. This flow has a speed of 1--3 km s$^{-1}$, so we adopt $v$ = 2 km s$^{-1}$. The EUV radiation, primarily from the direction of $\theta^1$ Ori C, ionizes the near surface of the sphere (the proplyd head)\footnote{The EUV photons also launch an ionized wind at $\sim$10 km s$^{-1}$ outward from the font. Because this wind is anisotropic, we focus our mass-loss calculation instead on the more spherically symmetric neutral outflow.}. This ``lights up" a portion of the outflow, allowing us to probe the gas density of the outflow at the ionization front as $\rho = n_e \mu m_H$, where $n_e$ is the electron density and $\mu$ = 1.35 is the atomic weight of the gas per electron. Here we compute $n_e$ to evaluate $\dot M$ in two ways: first, by invoking ionization equilibrium ($\dot M_\text{1}$) and second, by using the measured brightness of the free-free emission from the ALMA observations ($\dot M_\text{2}$).

The condition of photoionization equilibrium requires that
\begin{equation}
    \frac{\Phi}{4 \pi d_\perp^2} = \frac{\alpha_B n_e^2 r_\text{IF}}{3},
\end{equation}
where $\alpha_B$ = 2.6 $\times$ $10^{-13}$ cm$^3$ s$^{-1}$ is the recombination coefficient and $\Phi$ = 7.35 $\times$ 10$^{48}$ s$^{-1}$ is the flux of EUV photons from $\theta^1$ Ori C \citep{odell2017_ONCionization}. Solving this for $n_e$ and combining with the mass-loss relation above yields
\begin{equation}
\label{eq:Mdot1}
     \dot M_\text{1} = \mu m_H v (12 \pi \Phi / \alpha_B)^{1/2} d_\perp^{-1} r_\text{IF}^{3/2}. 
\end{equation}
Values of $\dot M_1$ for these proplyds are reported in Table \ref{tab:derived}. The mass-loss rates range from 0.6 to  3.1 $\times$ 10$^{-7}$ $M_\odot$ yr$^{-1}$, which are generally consistent with mass-loss rates inferred from prior studies of ONC proplyds \citep[e.g.,][]{storzer1999_proplyds}.

The second way we compute the mass-loss rate is by deriving $n_e$ from the brightness of the free-free emission. We do this via the ``emission measure," defined as $EM = n_e^2 L$, where $L$ is the line-of-sight path length through the ionized gas. We first compute the optical depth of the observed free-free emission as $\tau_\text{3.1mm} = I_\nu/B_\nu(T_e)$ (appropriate in the optically thin regime), where $I_\nu$ is the measured surface brightness of the ionization front in the Band 3 images and $B_\nu(T_e)$ is the Planck Function for an electron temperature of $T_e$ = $10^4$ K. We find $\tau_\text{3.1mm} \sim 10^{-3}$ in the ionization fronts of these proplyds. We then compute the emission measure following Equation (A.1b) of \citet{mezger1967_EM}:
\begin{equation}
    \left(\frac{EM}{\text{pc}~\text{cm}^{-6}}\right) = \left(\frac{\tau_\text{3.1mm}}{3.28 \times 10^{-7}}\right) \left(\frac{Te}{10^4\,K}\right) \left(\frac{\nu}{\text{GHz}}\right).
\end{equation}
We find $EM \sim 10^7$--$10^8$ pc cm$^{-6}$ for these sources.

We model the ionization front as a thin hemispheric shell of radius $r_\text{IF}$ and thickness $\Delta r$. The maximum line-of-sight path length through such a shell is $L = 2 \sqrt{2 r_\text{IF} \Delta r - \Delta r^2} \approx 2 \sqrt{2 r_\text{IF} \Delta r}$ when $\Delta r \ll r_\text{IF}$. The thickness of the ionization front is determined by the electron density as $\Delta r = (n_e \sigma)^{-1}$, where $\sigma = 6.3 \times 10^{-18}$ cm$^2$ is the ionization cross section. Thus, we can write the path length as $L = 2 \sqrt{2 r_\text{IF}/n_e \sigma}$. Substituting this into the definition of the emission measure and solving for $n_e$, we find
\begin{equation}
\label{eq:ne}
    n_e = \left(\frac{EM^2 \sigma}{8 r_\text{IF}}\right)^{1/3}.
\end{equation}
Using Equation \ref{eq:ne},  we find $n_e$ of 0.7--3 $\times$ $10^6$ cm$^{-3}$, generally consistent with estimates of $n_e$ from other methods \citep{bally1998_ONCHST}. $n_e$ = $10^6$ cm$^{-3}$ results in $\Delta r$ = 0.01 au, so the ionization fronts are indeed quite thin.

Combining Equation \ref{eq:ne} with the flow equation yields
\begin{equation}
\label{eq:Mdot2}
    \dot M_\text{2} = 2 \pi \mu m_H v \sigma^{1/3} EM^{2/3} r_\text{IF}^{5/3}. 
\end{equation}
We find $\dot M_\text{2}$ values of 3.2--18.4 $\times$ 10$^{-7}$ $M_\odot$ yr$^{-1}$, and we report the results in Table \ref{tab:derived}. 

The $\dot M_\text{2}$ calculation yields larger estimates for the mass-loss rates than $\dot M_\text{1}$ by factors of 1.5--9.0. There are several uncertainties inherent in both calculations. $\Phi$ in Equation \ref{eq:Mdot1} only accounts for ionizing photons from $\theta^1$ Ori C, whereas other massive stars in the cluster likely contribute EUV flux as well. Accounting for these would yield a larger $\dot M_\text{1}$. On the other hand, Equation \ref{eq:Mdot1} does not account for attenuation between the ionizing sources and the proplyds, and $d_\perp$ is smaller than the true distance to the sources, both of which would lead to an overestimated $\dot M_\text{1}$. Regarding $\dot M_\text{2}$, our use of the thin shell model for the ionization front differs from that used in prior studies. \citet{bally1998_ONCHST}, for example, used a more extended power-law profile. A thicker ionization front and correspondingly larger path length would yield lower values of $\dot M_\text{2}$ than we report. Overall, however, the decent agreement between two calculations that use different approaches gives confidence that these mass-loss rate estimates are accurate to within an order of magnitude or better.

\begin{figure}
\epsscale{1.17}
\plotone{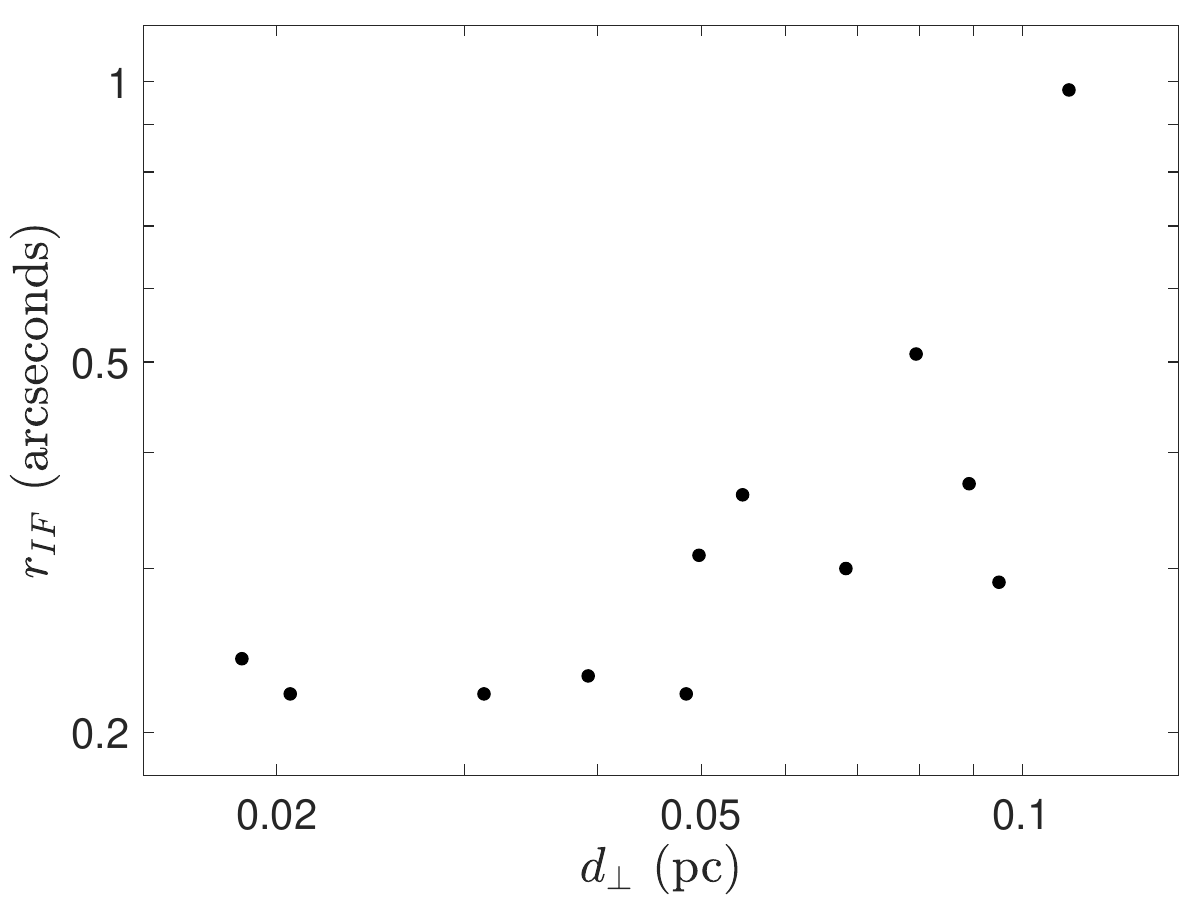}
\caption{Size of the proplyd ionization fronts versus their projected separation from $\theta^1$ Ori C.}
\label{fig:RIFDperp}
\end{figure}

Models by \citet{johnstone1998_photoevap} predict that if $\dot M$ is constant with $d_\perp$, we would observe $r_\text{IF} \propto d_\perp^{2/3}$. Figure \ref{fig:RIFDperp} illustrates the relation between these variables. While they are positively correlated---as borne out by the Spearman correlation (Table \ref{tab:spearman})---they do not appear to follow a power-law relation. Indeed, as shown in the calculations above, $\dot M$ is not constant; it varies by a factor of $\sim$5 within this sample.

We compute the photoevaporative timescale by dividing the disk gas mass (dust mass $\times$ 100) by the mass-loss rate. Considering we have two estimates of the disk dust masses and two estimates of the mass-loss rates, we compute both a minimum, $t_\text{min} = 100 M_d(T_d)/\dot M_\text{2}$, and maximum, $t_\text{max} = 100 M_d(20K)/\dot M_\text{1}$, estimate for the evaporative timescale. Values of $t_\text{min}$ range from 3.2 to 18.4 kyr and values of $t_\text{max}$ range from 37 to 1427 kyr; both are reported in Table \ref{tab:derived}. Most  of the timescales are shorter than the age of these systems (1--2 Myr), implying that such systems should no longer host disks. This discrepancy is referred to as the ``proplyd lifetime problem".

Solutions to the proplyd lifetime problem suggested in the literature include: (1) disk masses are larger than typically measured due to being optically thick at (sub)millimeter wavelengths \citep{clarke2007_photoevaporation}; (2) low-mass stars in the ONC are on radial orbits and thus spend relatively little time in the high-UV environment \citep{storzer1999_proplyds}; (3) multiple epochs of star formation occurred in the ONC, so some disks are younger than the median age \citep{beccari2017_ONCages}; (4) disks were shielded from the intense UV radiation by intervening gas in the cluster for a significant portion of their history \citep{qiao2022_diskstar,wilhelm2023_shielding}; or some combination of the above \citep{winter2019_lifetime}. Our results offer no new insights on arguments (2), (3), or (4). However, we find that the disks are optically thick so the dust masses may be significantly underestimated, emphasizing the importance of argument (1).

In the timescale calculations, we assumed a gas-to-dust mass ratio of 100, as is found in the ISM. Whether this ratio holds for protoplanetary disks is a matter of debate and is closely tied to uncertainties in the disk abundance of CO \citep[e.g.,][]{schwarz2018_COdepletion,zhang2019_CO}, the most readily observable gas tracer. The photoevaporation process preferentially removes gas and small grains, while the large grains, which we detect with ALMA, are not entrained in the outflow. Thus, photoevaporation could decrease the gas-to-dust ratio \citep{throop2005_evapplanetesimals}; although the large grains immune from evaporation are also the most susceptible to loss by inward drift in the absence of pressure traps \citep{sellek2020_dustphotoevap}. By modeling ALMA CO and HCO$^+$ observations of several ONC disks, \citet{boyden2023_ONCgas} found gas-to-dust ratios $\geq$ 100, suggesting our use of the ISM value is justified.

\section{Summary}
\label{sec:summary}

We present ALMA Band 3 (3.1 mm) images of twelve ONC proplyds. With high sensitivity and angular resolution, these observations detect and isolate dust emission from the protoplanetary disk and free-free emission from the surrounding ionization front. A comparison with images at both shorter (ALMA Band 7) and longer (VLA) wavelengths further solidifies the morphology of these systems. Our  main findings include:
\begin{itemize}
    \item The disks are generally small, probably due to truncation by external photoevaporation. The sizes are consistent with the broader population of ONC disks measured in previous surveys. The resulting planetary systems will likely be more compact than the solar system. The disk sizes correlate with the size of the ionization front and with the distance from $\theta^1$ Ori C, but disk size and disk flux are not correlated.
    \item The spectral indices measured from Band 7 to Band 3 are shallow ($\alpha \lesssim 2.1$) suggesting the dust emission is optically thick. High optical depth is expected for massive but compact disks.
    \item According to the standard calculation of disk dust mass from millimeter flux, these disks are massive enough to form planetary systems like our solar system despite ongoing external photoevaporation. The high optical depth of the dust emission means the true masses could be significantly higher.
    \item The photoevaporative mass-loss rates, calculated by assuming photoionization equilibrium and from the brightness of the free-free emission, are generally consistent with prior HST-based estimates and suggest photoevaporation timescales shorter than the $\sim$1 Myr age of ONC---underscoring the ``proplyd lifetime problem." Disk masses that are underestimated due to being optically thick are a potential solution to this problem.
\end{itemize}

Future observations at longer wavelengths with comparable resolution and sufficient sensitivity to detect the dust emission will further reveal the properties of ONC proplyds. Such measurements could come from deeper VLA observations, upcoming observations with ALMA Band 1, or future observations with the ngVLA. At these longer wavelengths, the optical depth of the dust decreases, providing more reliable disk masses and better measurements of the grain properties from spectral indices. The free-free emission is also more prominent at longer wavelengths. This may expand the sample of well-resolved proplyds to systems with fainter ionization fronts. Our work demonstrates that even at wavelengths where free-free emission dominates over the dust, the two can be reliably isolated with high-resolution observations.

\medskip
We thank the anonymous referee for many useful suggestions. We thank Andrew J. Winter for helpful discussions concerning the photoevaporative mass-loss rates. We thank NRAO staff, especially Brian Mason, for assistance with imaging the ALMA Band 3 data. We also gratefully acknowledge the use of NRAO computing facilities. This paper makes use of the following ALMA data: ADS/JAO.ALMA\#2015.1.00534.S and ADS/JAO.ALMA\#2018.1.01107.S. ALMA is a partnership of ESO (representing its member states), NSF (USA) and NINS (Japan), together with NRC (Canada), MOST and ASIAA (Taiwan), and KASI (Republic of Korea), in cooperation with the Republic of Chile. The Joint ALMA Observatory is operated by ESO, AUI/NRAO, and NAOJ. The National Radio Astronomy Observatory is a facility of the National Science Foundation operated under cooperative agreement by Associated Universities, Inc. N.P.B. acknowledges support from the Virginia Initiative on Cosmic Origins (VICO) and SOFIA Award 09-0181. L.I.C. acknowledges support from the David and Lucille Packard Foundation, Research Corporation for Science Advancement Cottrell Fellowship, NASA ATP 80NSSC20K0529, and NSF grant no. AST-2205698. T.J.H. acknowledges support from a Royal Society Dorothy Hodgkin Fellowship. A.G. acknowledges support from the NSF under AST 2008101 and CAREER 2142300. J.A.E and R.D.B acknowledge support by NSF AAG grant 1811290. J.A.E, J.S.K, and R.D.B acknowledge support by the National Aeronautics and Space Administration under Agreement No. 80NSSC21K0593 for the program ``Alien Earths." The results reported herein benefited from collaborations and/or information exchange within NASA’s Nexus for Exoplanet System Science (NExSS) research coordination network sponsored by NASA’s Science Mission Directorate.

\facilities{ALMA}
\software{CASA}

\bibliographystyle{aasjournal}
\bibliography{NPB}

\end{document}